# A spherical body in plasma simulated with adaptive mesh and algorithm refinement

Vladimir I. Kolobov[a*] and Robert R. Arslanbekov[b]

[a] The University of Alabama in Huntsville, Huntsville, AL 35899, USA

[b] CFD Research Corporation, Huntsville, AL 35806, USA

[*] Corresponding author: vik0001@uah.edu

## Abstract

An absorbing sphere in a weakly ionized plasma is the oldest kinetic problem of probe theory. It presents key difficulties in plasma simulation: disparate velocity scales, orbital motion, phase-space regions that no characteristic of the plasma reaches, a boundary set by a kinetic flux balance, and collisions that qualitatively change the answer. We use this problem to demonstrate the Plasma Adaptive Mesh and Algorithm Refinement (PAMAR) framework introduced in the companion paper [1]. Electrons and ions have separate solvers on their own adaptive $(r, v, \mu)$ octrees, or on reduced phase spaces at the angular-moment level (Fokker–Planck and two-stream models), coupled through a radial Poisson field solver. Two exact references, a method of characteristics for the collisionless problem and integration along the orbits for elastic scattering, audit the solvers. We compute the current–voltage characteristic of the probe on both electron and ion branches, and the profiles of the potential and of the moments of both species agree with the exact solution. Far from the sphere, the potential follows the Coulomb law in neither limit: it falls as $r^{-2}$ without collisions and as $r^{-1}$ only beyond the ion mean free path. With elastic electron scattering, the kinetic solver follows the exact dependence of the collected current on the mean free path, from free flight toward diffusion, to 1 percent. Finally, we demonstrate the ion model selection cell-by-cell, using a fluid model coupled to the 1d2v kinetic solver, with the interface set by the pressure anisotropy; with charge exchange, a sensor of ion trapping is also needed. According to exact solutions, the ion anisotropy extends to several probe radii, much farther than the electron loss cone.

## 1. Introduction

A body immersed in plasma acquires a charge controlled by the electron and ion currents it collects. Probe theory concerns the potential, the currents, and the perturbed plasma around it. In the classical works on this problem, Mott-Smith and Langmuir [2] gave the orbital-motion-limited (OML) currents, and Bernstein and Rabinowitz [3] and Laframboise [4] solved the collisionless Vlasov–Poisson problem for a sphere and a cylinder. Allen, Boyd and Reynolds [5] gave the radial-motion limit for cold ions, and Al'pert, Gurevich and Pitaevskii [6] treated bodies in the ionosphere. The same problem underlies dust-grain charging [7–9], the potential of small spacecraft, and measuring electron energy distributions with Langmuir probes [10–12].

Two features keep the problem relevant today. The first is ion trapping. For finite angular momentum, the attractive potential and the centrifugal barrier combine into a well, and orbits in that well touch neither the body nor the distant plasma. A stationary collisionless theory cannot populate them, since collisions and the potential's history set their occupancy. Laframboise assumes they are empty [4], with the warning that his results may be of limited use where they predict potential wells. Charge-exchange collisions fill those orbits in any real discharge, and the trapped ions change the screening, the collected current and the floating potential [7,8]. Stationary backward-Liouville solvers face the same limit numerically [13]. The second is collisions of electrons. The Druyvesteyn formula, which recovers the electron energy distribution from the second derivative of the probe current, rests on collisionless electron trajectories inside the perturbed region. When the electron mean free path becomes comparable to the probe size, the current is reduced in an energy-dependent way; in the opposite limit, the first derivative carries the distribution instead [11,12,14]. Both effects need a kinetic description with angular momentum and a collision operator, in a geometry where the two species see incommensurate phase spaces.

This is the setting for which Adaptive Mesh and Algorithm Refinement (AMAR) was conceived [15–17], where each species and each region of phase space should be computed on its own adaptive mesh with the cheapest adequate model. The companion paper [1] introduces PAMAR, the plasma extension of AMAR, and demonstrates it on a capacitively coupled discharge and, briefly, on the floating sphere. In the present paper, we treat the sphere in more detail. Section 2 states the equations and collision operators, Section 3 presents the hierarchy of transport models and the corresponding species solvers, and Section 4 presents the Poisson field solver and boundary conditions. Section 5 presents the exact solution for the collisionless case and the ion-trapping problem. Section 6 compares the electron and ion solvers on common benchmarks, and Section 7 computes the collisionless current–voltage characteristic (CVC) and the far-field of the potential. Section 8 turns to elastic electron scattering and its effect on the collected current; the consequences for probe diagnostics are left to a separate paper. In Section 9, we select the ion solver cell by cell using a fluid–kinetic split with the 1d2v octree. Its electron counterpart (the 1d2v solver confined to the loss-cone layer with a Fokker–Planck model outside the cone) is reported in [1]. Section 10 summarizes.

## 2. Basic equations

### 2.1. Kinetic equation in spherical phase space

We consider a perfectly absorbing sphere of radius $R_p$ in an unmagnetized plasma whose electrons and ions are Maxwellian far away, with density $n_\infty$ and temperatures $T_e$ and $T_i$. Spherical symmetry without a magnetic field leaves the radius $r$, the speed $v$ and the cosine $\mu = v_r/v$ of the angle between the velocity and the radius as independent variables. This 1d2v phase space is the smallest that still contains angular momentum. The distribution function $f_s(r, v, \mu, t)$ of species $s$ with charge $q_s$ and mass $m_s$ obeys

$$\frac{\partial f}{\partial t} + v\mu\frac{\partial f}{\partial r} + \mu a\frac{\partial f}{\partial v} + (1-\mu^2)\left(\frac{v}{r} + \frac{a}{v}\right)\frac{\partial f}{\partial \mu} = C_s[f], \quad (1)$$

with $a = q_s E_r/m_s$. In the solvers, we write it for the conservative variable $Y = f v^2 r^2$, so that the finite-volume update conserves particles exactly on the adaptive octree. Two invariants of the collisionless motion, the energy and the specific angular momentum,

$$\varepsilon = \frac{m_s v^2}{2} + q_s\phi(r), \quad \ell = rv\sqrt{1-\mu^2}, \tag{2}$$

combine into the effective potential $U_\ell(r) = q_s\phi(r) + m_s\ell^2/2r^2$. The pitch-angle coefficient of Eq. (1) vanishes on the turning-point envelope $v^2r^2 =$ const of $U_\ell$, and its sign decides whether a group of orbits is focused or defocused by the combined centrifugal and electrostatic forces [18].

## 2.2. Collisions: elastic scattering of electrons and charge exchange of ions

Electrons scatter elastically on the atoms. The energy exchanged is of order $m/M$ and is neglected on the scale of the probe, so scattering conserves an electron's speed and energy. Isotropic scattering relaxes $f$ toward its angular mean at fixed $(r, v)$ (the Lorentz model),

$$C_e[f] = \nu_m(v)(f_0 - f), \quad f_0 = \frac{1}{2}\int_{-1}^{1} f d\mu, \tag{3}$$

with $\nu_m = Nv\sigma_m$. We quote it through the mean free path $\lambda = v_T/\nu_m$ of an electron of speed $v_T = (2T_e/m)^{1/2}$; the benchmarks of Section 8 use a constant $\nu_m$, for which the free path of an electron grows with its speed. Because the speed is conserved, every energy group is an independent problem, which makes the exact reference of Section 3 possible. Anisotropic scattering replaces $\nu_m$ by the momentum-transfer frequency in the first angular moment.

Ions undergo resonant charge exchange (CX) with their parent atoms. We use the BGK form throughout, in which an ion is replaced by a slow one born from the atom distribution,

$$C_i[f] = -\nu_{CX}(v)f + S_{CX}(r)\varphi_M(v), \quad S_{CX} = \int \nu_{CX} f\, d^3v, \quad \nu_{CX}(v) = \frac{v_{th,i} + v}{\lambda_i}, \tag{4}$$

where $\varphi_M$ is the normalized Maxwellian of the gas temperature, $S_{CX}(r)$ is the local rate at which ions are lost by CX, so that the operator conserves ions in every shell, and $\lambda_i = 1/N\sigma_{CX}$ is a velocity-independent mean free path. The charge exchange has two effects that matter here. It removes fast ions and moves ions from passing orbits onto trapped orbits. The second effect is singular, since at any finite $\lambda_i$ the trapped population is set by a balance of filling and draining that has no collisionless counterpart [6].

## 2.3. Poisson's equation and the dimensionless parameters

The potential obeys the radial Poisson equation

$$\frac{1}{r^2}\frac{d}{dr}\left(r^2\frac{d\phi}{dr}\right) = -4\pi e(n_i - n_e) \tag{5}$$

with $\phi(R_p) = \phi_p$ and $\phi \to 0$ far away. Temperatures are measured in energy units (electronvolts) throughout, so the Boltzmann constant does not appear. Measured in its own temperature, each species sees only the drop $\chi_p = -e\phi_p/T_e$, the ratio $R_p/\lambda_D$ of the probe radius to the Debye length, the temperature ratio $T_e/T_i$, and, with collisions, $\lambda/R_p$ and $\lambda_i/R_p$. The mass ratio does not appear in the steady collisionless problem at all, since the distribution and every density depend on $q_s\phi/T_s$ only, and $M/m$ enters through the fluxes, i.e. only where the electron and ion currents are compared [4,6]. This fact, used in Sections 4.3 and 7.4, allows using an artificial mass ratio at a prescribed bias and recovering the physical one afterward. Currents are given as $i_s$, the collected current in units of the random current $4\pi R_p^2 n_\infty \overline{v}_s/4$ of the same species, and the code units are listed in [1]. Note that the solver's input screening parameter and the electron Debye length differ by a constant factor (0.424).

## 3. The hierarchy of transport models and the species solvers

A hierarchy of models describes each charged species (Figure 1 of [1]). For the sphere, the hierarchy has three levels, each realized in PAMAR by one or more species solvers: full kinetics in the 1d2v phase space; the angular-moment level, where a few angular moments with a closure replace the dependence on the pitch angle and the full dependence on the speed is kept [1]; and the fluid level. Three angular-moment models are used here: the Lorentz model, which leads to the Fokker–Planck equation for the isotropic part of the distribution, and the two-stream models 1d1v(FB) and 1d1v(S). Every species solver exports the same six functions to the field solver: advance to a given time, stop, return the density and flux at a point, return the stable time step, and fill the field solver's shell profiles of density and flux in one sweep of the species solver's own mesh. The field solver therefore remains unchanged with respect to any of them. Table 1 lists the species solvers and the two exact references used to audit them.

**Full kinetics, 1d2v.** Equation (1) is solved on an adaptive octree in $(r, v, \mu)$ with conservative advection of $Y$, the face velocities $(v\mu, a\mu, (1-\mu^2)(v/r + a/v))$, and the collision operators (3) or (4). The radial axis can be a mapped coordinate graded toward the surface (Section 4.2). This solver uses no orbit information and takes any collision operator and any time dependence at the expense of computing time (the collisionless steady state takes 18 min, 3.7 h and 54 h on one core on $32^3$, $64^3$ and $128^3$ meshes).

**Angular moments, Lorentz model: Fokker–Planck, 1d1ε.** Elastic scattering makes the electrons nearly isotropic outside the loss cone, so that the first two angular moments suffice (the Lorentz approximation). The isotropic part $f_0(r,\varepsilon)$ at fixed total energy $\varepsilon$ then diffuses along the radius; in the steady state, $\partial_r[r^2(v^3/3v_m)\,\partial_r f_0] = 0$ with $v = [2(\varepsilon - U(r))/m]^{1/2}$ and $U$ the potential energy (the nonlocal approach of [19,20]). The boundary conditions are $f_0 = f_\infty(\varepsilon)$ at the outer boundary and the half-range (Marshak) condition $f_0 = (2v/3v_m)\,\partial_r f_0$ at an absorbing surface. The model is exact for the electrons trapped below the wall barrier and cannot represent the empty loss cone (Section 6.3).

**Angular moments, two-stream model: 1d1v(FB).** The simplest two-stream model is the forward–backward closure $f = f_+\delta(\mu-1) + f_-\delta(\mu+1)$, which places every particle on a radial line. For ions, it is the Allen–Boyd–Reynolds radial-motion limit [5], realized on an adaptive quadtree in $(r, v_r)$. It carries no angular momentum and therefore cannot represent orbital collection or trapping; it is exact for ions born at rest in a monotone potential, which is why it serves well in the planar discharge of [1] and poorly around a sphere fed from a reservoir.

**Angular moments, two-cone model: 1d1v(S).** Charge exchange does not make the ions isotropic: they move along orbits with angular momentum, which a two-stream closure on orbit cones follows. The reduced model keeps $(r,\varepsilon)$ and replaces the dependence on $\mu$ by a step function whose steps follow the orbit geometry, in the spirit of Svettsov's method for spherical radiative transfer [21]. With $A(r) = r^2v^2$ at the given energy, $a(r) = \min_{s\geq r} A(s)$ is the largest $\ell^2$ still connected to the outer boundary and $b(r) = \min_{s\leq r} A(s)$ the largest connected to the surface; the regions are the arriving orbits ($\mu < 0$, $\ell^2 \leq a$), the reflected band ($\mu > 0$, $b < \ell^2 \leq a$), the loss cone ($\mu > 0$, $\ell^2 \leq b$) and the bound orbits ($\ell^2 > max(a,b)$). The edges are made of orbits, so no streaming flux crosses them, and integrating the kinetic equation over the regions gives

$$\frac{d}{dr}\left[\frac{a}{2}f_-\right] = -\frac{f_-}{2}\frac{da}{dr} + \nu r^2 v\, w_-(\langle f\rangle - f_-),$$
$$\frac{d}{dr}\left[\frac{a-b}{2}f_+\right] = \frac{f_-}{2}\frac{da}{dr} + \nu r^2 v\, w_+(\langle f\rangle - f_+), \qquad (6)$$
$$\frac{d}{dr}\left[\frac{b}{2}f_c\right] = \nu r^2 v\, w_c(\langle f\rangle - f_c),$$

where $w_-$, $w_+$, $w_c$ are the measures of the regions in $\mu$, $f_- = f_\infty(\varepsilon)$ at the outer boundary and $f_c = 0$ at the surface; the sum of the three is the conservation of $r^2\Gamma$. Without collisions, the solution $f_\pm = f_\infty$, $f_c = 0$ is exact; with collisions the uniform step is a closure. A fourth region carries, for an attracted species in a bounded domain, the orbits that leave the surface and return to it. As an ion solver, the model also carries the occupancy of the bound region as an explicit variable, a trapped bath fed and drained by charge exchange, which a stationary orbit theory leaves undetermined.

**Fluid, 1d0v.** We use the two-moment model with the ion temperature held at the gas temperature at every radius, $p = nT_i$,

$$\partial_t n + \frac{1}{r^2}\partial_r(r^2 n u) = 0, \qquad (7)$$

$$Mn(\partial_t u + u\partial_r u) = qnE - \partial_r p - Mn\nu_{CX}u, \qquad (8)$$

with the drag form of the same charge-exchange rate. It is blind to orbital physics by construction and meets the sonic-point difficulty of the radial-motion limit [22]; dropping inertia gives the drift–diffusion model. For the electrons, the corresponding level is the Boltzmann relation.

**Exact references.** Two semi-analytic solvers serve as references for the species solvers. Both are stationary, and each holds for one collision model, so that they audit the PAMAR species solvers and do not replace them: the 1d2v solver is time-dependent, takes any collision operator and does not require integrable orbits. The method of characteristics (Section 5) is exact for the collisionless problem with any potential profile. The integration along the orbits is exact for elastic scattering, since at fixed energy and angular momentum the kinetic equation is an ordinary differential equation along the orbit, and scattering acts only through the isotropic part,

$$f(\tau) = f(0)e^{-\tau} + \int_0^\tau \langle f\rangle\, e^{-(\tau-\tau')}d\tau', \quad d\tau = \frac{\nu_m r\, dr}{\sqrt{A(r) - \ell^2}},$$
$$\langle f\rangle = \frac{1}{4\sqrt{A}}\int_0^A \frac{f_+ + f_-}{\sqrt{A - \ell^2}}\, d\ell^2, \qquad (9)$$

with $f_-$ started from the Maxwellian at the outer boundary, $f_+ = f_-$ at the turning point and $f_+ = 0$ at the surface for the orbits that reach it. With $\langle f\rangle$ constant in each radial cell, $\ell^2$ nodes placed between the values of $A$ at the cell faces so that every turning point falls inside a cell, and the square-root singularities integrated analytically, the problem for $\langle f\rangle$ is a dense linear system per energy, solved directly. Its restriction is that $A(r)$ must increase outward, which holds for a repelled species and fails for an attracted one at low energy. A further stationary solver of this kind serves the ions with charge exchange in the far field (Section 7.6). It integrates Eq. (1) with the operator (4) along the orbits on a lattice of energy and angular momentum, bound orbits included, and finds the birth rate $S_{CX}(r)$ together with the potential by Newton iteration; the newborn ions are weighted by $\nu_{CX}(v)$, so that the Maxwellian is an exact equilibrium at any mean free path.

**Table 1.** The models for one species around the sphere. Costs are for the benchmark of Section 6, collisionless except in the orbit and FP rows (elastic scattering); the 1d2v times are for levels 5, 6 and 7.

| model | phase space | angular momentum | collisions | role here | cost (one core) |
|---|---|---|---|---|---|
| MoC (reference) | quadrature over $\ell^2$ | exact | none | collisionless anchor, warm start, audit | 0.2 s |
| orbits (reference) | $\langle f\rangle(r,\varepsilon)$ | exact | elastic | collisional anchor | 3–20 s |
| 1d2v kinetic | $(r,v,\mu)$ octree | yes | any | electron and ion solver | 18 min – 54 h |
| 1d1v(S) | 3–4 regions in $(r,\varepsilon)$ | via barriers | elastic, CX (closure) | electron and ion solver | 0.4 s |
| FP 1d1ε | $f_0(r,\varepsilon)$ | Lorentz model | elastic | electron solver outside the loss-cone layer | 0.4 s |
| 1d1v(FB) | $(r,v_r)$ quadtree | no | CX | ion solver | minutes |
| fluid 1d0v | $r$ | no | CX drag | ion solver | seconds |

# 4. The field solver, boundary conditions and the floating potential

## 4.1. Poisson's equation on Laframboise's net

We solve Eq. (5) by a Thomas algorithm on radial shells. The species moments are gathered from each species solver as conservative finite-volume sums over its leaf cells, each cell's content distributed over the shells it overlaps with the weight of its $r^2dr$ volume. The shells follow Laframboise's computation net [4], uniform in a grid coordinate that is logarithmic in $r$ for small $R_p/\lambda_D$, uniform in $1/r$ for intermediate values and stretched by a power law beyond. The outer boundary carries the power-law condition $d\phi/dr = -2\phi/r$, which allows the domain to end a few Debye lengths from the sheath. The field solver provides the potential $\phi_B$ at the outer boundary, which shifts the reservoir Maxwellians of the species solvers (Section 4.2). The shells are independent of the radial meshes of the species solvers. Those are graded toward the surface, where the loss cone opens (Section 6.1), while the field needs the Debye length resolved through the sheath and the presheath out to the outer boundary, and the conservative transfer of the moments lets each mesh serve its own purpose. The species meshes can also be graded in $\mu$, toward the edge of the loss cone, which we leave to a forthcoming paper.

## 4.2. The surface and the reservoir

Both species are absorbed at $r = R_p$, and a Maxwellian reservoir supplies them at $r = R_{out}$. We impose the exact fluxes there: the wall emits nothing, and the reservoir injects the Liouville state from infinity, $v\mu f_\infty(\varepsilon)$ with $\varepsilon$ including $q_s\phi_B$. A third condition closes the $v = 0$ face, through which decelerated particles otherwise leave phase space at their turning points instead of turning through $\mu$. With the three conditions, the net radial current is the same through every face of the octree to $2 \times 10^{-4}$, and the only error left is the extrapolation of the cell-average distribution to the wall face, which is first order in the size of the wall cell. A radial map graded toward the surface, with the Jacobian carried in the conserved variable, reduces this error at no extra cells (at level 5 the wall cell is that of a uniform level-7 mesh, at one sixty-fourth of the cells) [1].

### 4.3. The floating potential

The wall potential either floats by the current-balance feedback $d\phi_p/dt = GI_{net}$, the discrete analog of wall charging, or is held at a prescribed bias. The second option is preferable, since at a fixed bias the profiles are physical and mesh errors appear directly in the currents instead of being absorbed by a shift in a feedback variable. In addition, the steady state at fixed $\chi_p$ does not depend on the mass ratio (Section 2.3), and the floating state is recovered from the computed characteristic and the balance

$$i_e(\chi_f) = i_i(\chi_f)\sqrt{\frac{mT_i}{MT_e}}. \tag{10}$$

For the same reason, the two species need not advance on a common clock, and each species solver takes its own CFL step per field iteration [1]. At $M/m = 100$ and $T_i = T_e$, the ions take 4.2 steps of their own per electron step, and the ion state reaches a given time 3.1 times faster in computing time than with a common clock. Both shortcuts hold only for the steady state, and only where it is unique. With trapped orbits it is not unique, and Section 5.3 explains why the trapped population must then be computed on the physical collisional clock. The floating runs on the large domain therefore advance both species in lock step (Section 9.4).

## 5. The collisionless reference and the problem of ion trapping

### 5.1. The method of characteristics

In a stationary collisionless plasma, $f$ is constant along every orbit (Liouville's theorem). A point in phase space therefore carries the reservoir value $f_\infty(\varepsilon)$ if its orbit reaches back to the outer boundary without striking the sphere, and zero otherwise. An orbit at radius $r$ is blocked by the higher of the barriers on either side of it, so with the two barrier functions

$$B_{out}(r,\ell^2) = max_{s\geq r} U_\ell(s), \quad B_{in}(r,\ell^2) = max_{R_p\leq s\leq r} U_\ell(s), \tag{11}$$

the occupancy rule is: inbound points ($\mu < 0$) are populated if $\varepsilon \geq B_{out}$, the orbit having cleared the outer barrier on its way in; outbound points are populated if $B_{out} \leq \varepsilon < B_{in}$, the orbit having been turned by the centrifugal barrier before it reached the surface; everything else is empty, namely the loss cone, which holds the outgoing legs of absorbed orbits, and the bound orbits, which no orbit from the plasma reaches. With the occupancy known, the angular integrals collapse, and the energy integrals evaluate in closed form through incomplete gamma functions. The only numerics are a smooth quadrature over $\ell^2$ and, when the field is solved with the densities, a Poisson iteration. There is no velocity mesh and no statistical noise, and one species evaluation takes a fraction of a second.

We tested the solver against four independent gates, which it passes. At $\phi = 0$ the sphere casts a pure geometric shadow in velocity space, and the density is Al'pert's free-molecular profile

$$\frac{n(r)}{n_\infty} = \frac{1}{2}\left[1 + \sqrt{1 - \frac{R_p^2}{r^2}}\right], \quad i = 1, \tag{12}$$

which is exactly half the ambient density at the surface, because the outgoing hemisphere is empty there. The MoC reproduces it to rounding, and the orbit solver and 1d1v(S) to $3 \times 10^{-5}$ and $5 \times 10^{-5}$ of $n_\infty$ (Figure 1(a)). Self-consistently coupled to Poisson's equation it reproduces Laframboise's equal-temperature Table 5c to $1.4 \times 10^{-4}$, the cold-ion block of his Table 5b ($T_i/T_e = 0.1$, $R_p/\lambda_D = 10$) in all eleven rows within 1 percent (Section 7.3), and the OML ceiling $i = 1 + |\chi_p|$ to $3 \times 10^{-4}$ in the Coulomb limit; $r^2\Gamma$ is constant to $4 \times 10^{-16}$. Table 5c is reproduced to 0.7 percent using 1d1v(S) to provide the attracted-species density inside the same Poisson iteration, with no velocity mesh. A second orbit solver, with exact orbit integrals and a Newton field solve, reproduces Table 5c to 0.02 percent and agrees with the MoC to $1.3 \times 10^{-4}$ in both currents at the twelve biases of Section 7.5 ($R_p/\lambda_{De} = 8$, $T_i = T_e$).

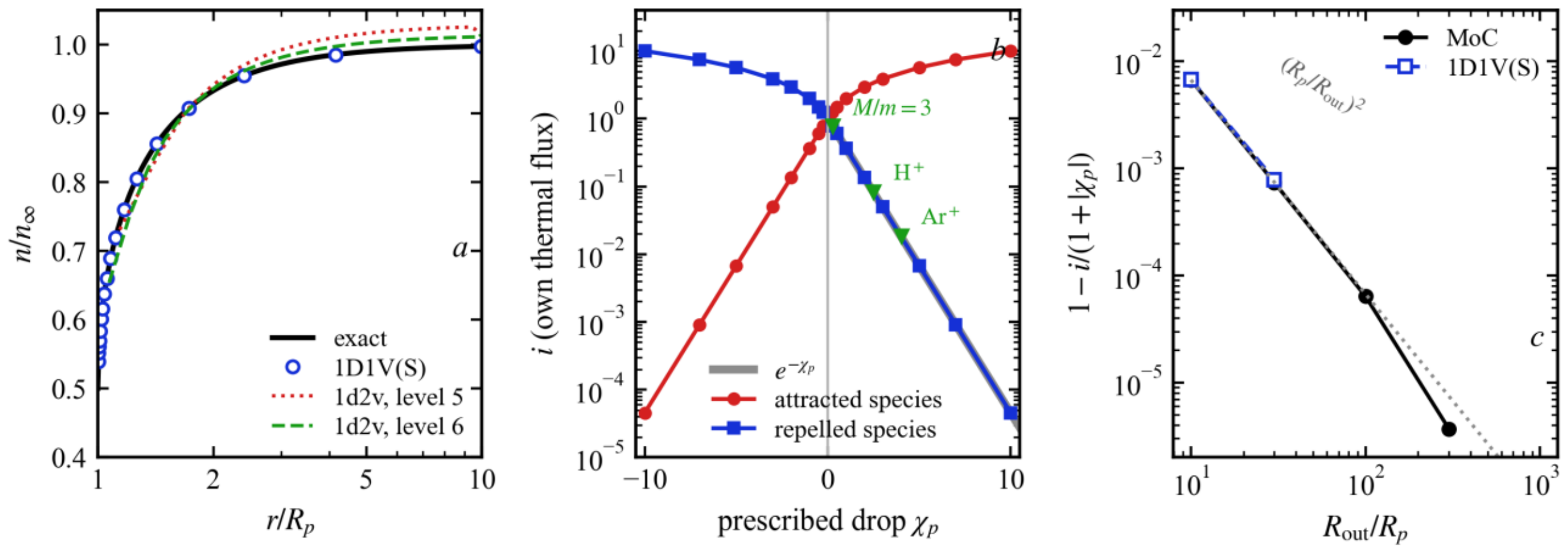

**Figure 1.** The collisionless gates. (a) The field-free density: Eq. (12) (line), 1d1v(S) (circles) and the 1d2v solver in its first form at levels 5 and 6 (dotted, dashed; the corrected solver lies on the line to 3 percent at the surface). (b) Both currents at a prescribed drop, $R_p/\lambda_D = 1$, $T_+/T_- = 1$, each in its own thermal flux; grey: $e^{-\chi_p}$, which the repelled branch follows exactly; triangles: the floating points for three mass ratios. (c) Approach to the OML limit as the outer boundary is moved out, $\chi_p = -2$.

## 5.2. Orbit groups in phase space

The barrier functions sort every point of the $(v, \mu)$ plane at a given radius into four groups (Figure 2). For the repelled electrons of the benchmark of Section 6 ($\chi_p = 2$, Coulomb potential vanishing at $R_{out} = 10R_p$) at $r = 2R_p$ the plane is 50 percent arriving, 47 percent reflected and 3 percent loss cone. There are no bound orbits, since a monotone repulsive potential has no minimum of $U_\ell$ at any $\ell$, so that the repelled species cannot be trapped. For the attracted species, the four regions hold 35, 28, 31, and 6 percent. In a screened self-consistent potential, the bound region grows, and at $\chi_p = 5$, $\lambda_{De} = 2R_p$ and $T_i = T_e$ the effective potential has a well and a barrier together (an O-point inside an X-point) for $\ell^2 = 7.2$–$14.6\, R_p^2 T_i/M$, with O-points at 1.0–2.7 $R_p$ and X-points at 3.8–7.3 $R_p$. At $r = 2R_p$ the bound region is then 45.6 percent of the plane against 27.2 percent arriving, 19.2 percent reflected and 8.0 percent loss cone. The edge of the loss cone $\mu_c = [1 - A(R_p)/A(r)]^{1/2}$ for a repelled species approaches the geometric shadow $\left(1 - R_p^2/r^2\right)^{1/2}$ at high energy (Figure 2(c)); its solid angle is the quantity that decides, in Section 9, where a species needs a kinetic description.

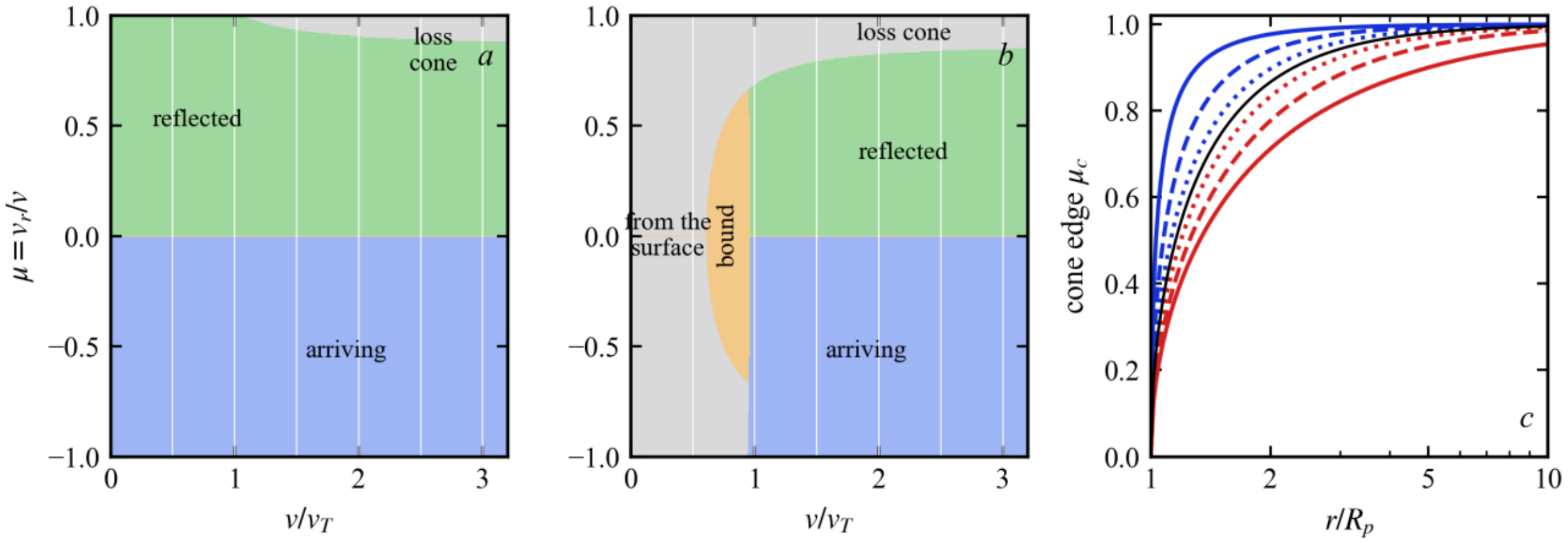

**Figure 2.** Occupancy of the $(v, \mu)$ plane at $r = 2R_p$ for repelled (a) and attracted (b, $\chi_p = -2$) electrons in the Coulomb benchmark, $|\chi_p| = 2$: arriving orbits (blue), reflected orbits (green), orbits that begin on the surface (grey: the loss cone and, for the attracted species below the escape speed, its returning legs) and bound orbits (orange). Without collisions $f = f_\infty(\varepsilon)$ in the blue and green regions and zero elsewhere; the white lines are contours of $f$. (c) Edge of the loss cone against radius for energies 0.25, 1 and 4 $T_e$ above the threshold (solid, dashed, dotted), repelled (blue) and attracted (red); black: the geometric shadow $\left(1 - R_p^2/r^2\right)^{1/2}$.

## 5.3. The ion hole

The bound region touches no boundary, so a stationary collisionless solver cannot populate it: tracing characteristics back to the plasma never reaches those orbits, and the problem as posed does not determine their occupancy. Al'pert and co-workers state that particles on finite orbits are not connected to the conditions at infinity and that collisions, however rare, determine their number [6]. Laframboise assumes them empty [4], and we do the same in the MoC. This convention matters, since in the screened example above the empty region is nearly half of the accessible plane (Figure 3). Chiabó and Sánchez-Arriaga [13] analyzed the practical cost of stationary backward-Liouville solvers. According to their analysis, the trapped/passing classification is itself a numerical act; trapped particles are the population most damaged by field-reconstruction errors, and for non-integrable configurations the boundary becomes fractal. Their recommendation for integrable problems, to compute phase-space boundaries rather than the distribution at every mesh point, is the basis for 1d1v(S). The time-dependent 1d2v solver needs no such classification: it follows the distribution through the transient with an explicit collision operator, so that the occupancy of the trapped orbits is computed rather than assumed.

Figure 3 shows the radial evolution of the ion distribution (the four slices) and the orbits that produce it (the characteristics) for the screened example of Section 5.2, both computed from the same converged state. One can see that the ion distribution in $\mu$ stays broad all the way to the surface despite strong focusing (blue lines), and the two observations do not contradict each other. Focusing is the $1/r^2$ convergence plus acceleration that raises the ion density near the sphere; the angle is governed by angular-momentum conservation, $\ell = bv_\infty$ for an ion with impact parameter $b$. At radius $r$ its pitch satisfies $\sin\theta = \ell/(rv)$: the transverse speed is amplified inward as $1/r$ by very nearly the factor by which acceleration raises the radial speed, and the two effects cancel. Inside the capture radius every impact parameter up to $b_{max} = R_p\sqrt{1 + \chi_p T_e/T_i}$ is collected, those impact parameters map onto the entire inward hemisphere, and Liouville's theorem keeps $f$ uniform across that band, $f = f_\infty(\varepsilon)$ independent of $\ell$. The reflected orbits (green), turned by the centrifugal barrier, fill the outbound side as well. The broad $\mu$ distribution is therefore the signature of orbital collection itself, carried entirely by the passing ions. Trapped ions are a separate population, since a trapped orbit oscillates between two turning points where $v_r = 0$, so that its $\mu$ swings symmetrically through zero and never reaches $\pm 1$. In the collisionless state of Figure 3, that family is empty, but its characteristics exist, and the red loops show where charge-exchange-fed ions would accumulate (closed curves at $r \approx 1 - 3R_p$, $v/v_0 \approx 0.6$–$0.9$, concentrated near $\mu \approx 0$).

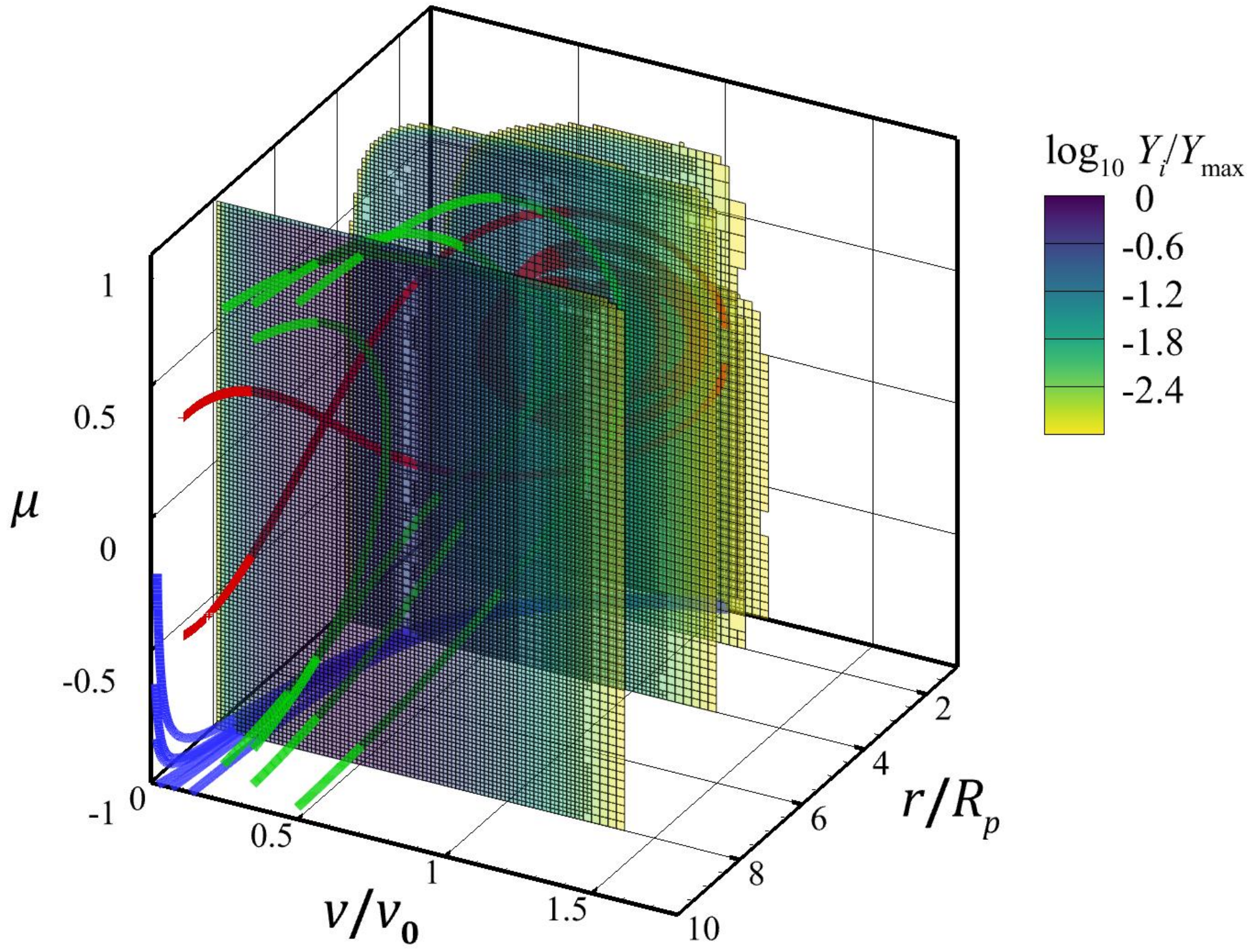

**Figure 3.** Ion phase space around the sphere at a prescribed bias (the screened example of Section 5.2, $\chi_p = 5$, $\lambda_{De} = 2R_p$; potential from the self-consistent MoC solution). Velocity in units of the sheath-fall speed $v_0$, $Mv_0^2/2 = e|\phi_p|$: an ion falling from rest through the full wall potential arrives at the sphere at $v/v_0 = 1$. Translucent slices: the ion distribution at four radii from the cold quasi-neutral bulk to the focused, sheath-accelerated beam at the sphere. Streamlines are colored by orbit class from the exact collisionless classification of the same potential: blue, arriving orbits that reach the surface; green, orbits reflected by the centrifugal barrier; red closed curves, trapped orbits inside the X/O island, empty in this collisionless state and the population that a charge-exchange source would fill; red dashed, the separatrix through the X-point. The loops drawn have their O-points at 1.3–2.0 $R_p$ and their X-points at 4.8–6.6 $R_p$ ($l^2 = 9$–$12.5\ R_p^2 T_i/M$), inside the island of Section 5.2.

Two processes outside a stationary collisionless model fill the hole in a real plasma. The first is collisions, since a charge-exchange event moves an ion from a passing orbit onto a bound one. The second is the transient in which the potential well builds, during which a passing ion can become bound as the well deepens. Both are present in PAMAR, since the kinetic solvers are transient with an explicit charge-exchange operator, so their trapped population emerges from the self-consistent history, and 1d1v(S) promotes hole occupancy as a dynamical variable fed and drained by collisions. For this reason, we start every run in this paper from the exact collisionless state with the bound orbits empty, so that whatever appears afterward is put there by the collision operator, not by the initial condition. Started this way, the kinetic ion solver at fixed bias shows the filled hole in the field (Figure 4). The same run marched without collisions drifts slowly in the attracted current, because numerical diffusion in velocity fills the bound region from the passing orbits (Section 6.1). On a mesh, the hole is thus filled by the scheme unless physics fills it faster.

Figure 4 shows the kinetic $\lambda_i$ sweep at fixed bias ($\chi_p = 5$, $R_p/\lambda_{De} = 8$, $T_i = T_e$, $R_{out} = 4.5R_p$), in which the trapped population appears in the field. The currents of this sweep are quoted in [1]. The ratio $R_p/\lambda_{De} = 8$ is that of the screened case in [1], where the Debye length was set through the code's Poisson coupling; we keep it so the runs can be compared directly with those of [1]. The collected ion current falls 1.96 times between the collisionless and the $\lambda_i = R_p$ runs, as charge exchange removes the sheath-accelerated ions, and at $\lambda_i = 3R_p$ and $R_p$ $\chi(r)$ outside the sheath falls below zero, i.e. the potential rises above the reservoir level (Figure 4(b)). This inversion is due to the space charge of the charge-exchange-born ions accumulating on bound orbits in the presheath (the filled ion hole), and it is of the order of a few hundredths of $T_e$. The comparison with 1d1v(S), whose trapped bath is the variable that this space charge measures, is made in Section 6.4. In a planar discharge, the ions are born at rest inside the plasma by ionization and acquire no angular momentum in charge exchange, so that a radial description is exact and no hole arises. This is why fluid ions fare better there than around the sphere, where they over-collect by 50–69 percent (Section 6.4). Around the sphere, the reservoir supplies ions with a distribution of impact parameters, and 1d1v(S) is needed exactly where impact parameters exist to be distributed.

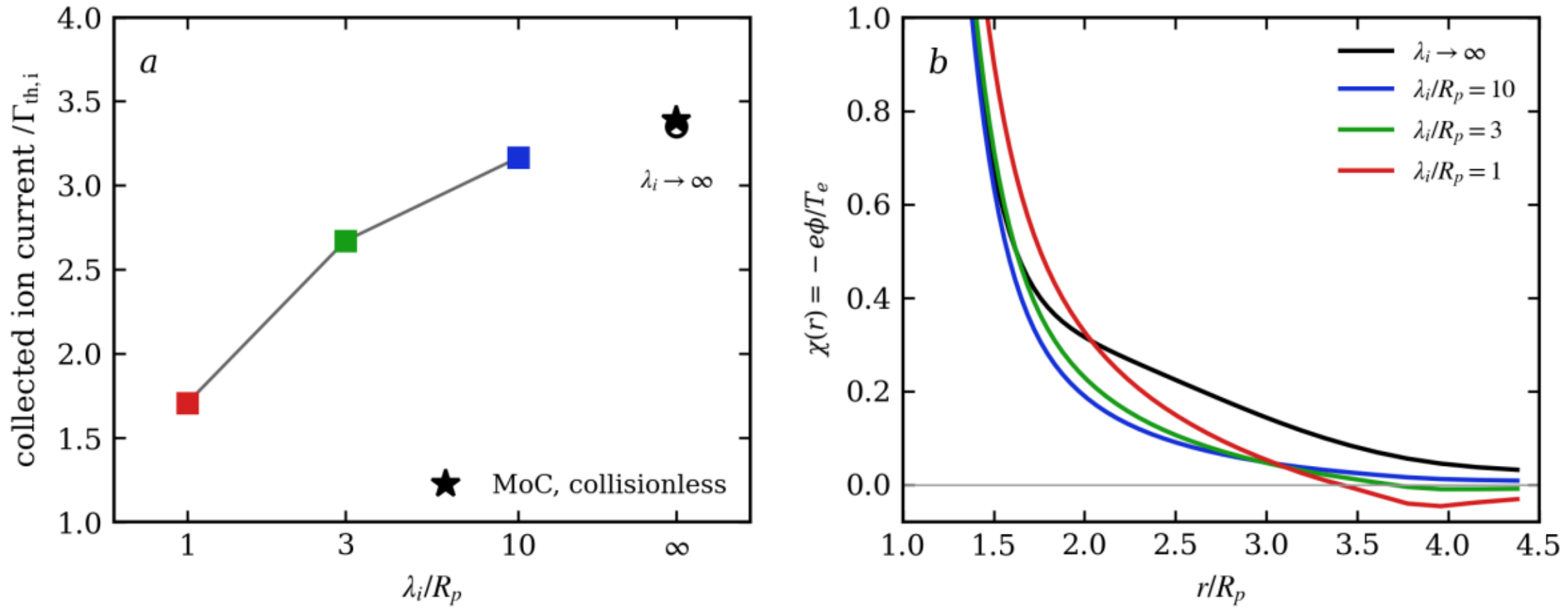

**Figure 4.** The kinetic $\lambda_i$ sweep at fixed bias ($\chi_p = 5$, $R_p/\lambda_{De} = 8$, $T_i = T_e$, $R_{out} = 4.5R_p$ with the $r^{-2}$ tail): (a) collected ion current at the wall face against the ion mean free path, with the MoC collisionless value (star) and the collisionless run (open circle); (b) the potential $\chi(r) = -e\phi/T_e$ for the four runs on a linear scale that cuts the sheath ($\chi_p = 5$ at the wall) to show the presheath and the far field: with charge exchange $\chi$ drops below zero (the potential rises above the reservoir level) beyond $r \approx 3.4R_p$, the space charge of the ions born on bound orbits.

## 5.4. The non-monotonic attracted density

Angular momentum produces a second, smaller feature that a radial model cannot. Just outside the surface, the outbound hemisphere is strictly empty, since everything that arrives is absorbed, so the attracted density there comes from inbound orbits alone. A fraction of a Debye length out, the reflected group begins to fill the outbound side, rising from exactly zero, while the inbound part falls under acceleration and geometric dilution; where the rise beats the fall, the density has a local maximum (Figure 5). Beyond the maximum, the density falls to a minimum about $3\lambda_D$ from the surface and rises again toward the quasi-neutral plasma. Cooling the ions suppresses the maximum (4.9, 1.9, and 1.3 percent at $T_i/T_e = 0.5$, 0.25 and 0.2), because cold ions fall into nearly radial orbits and are absorbed rather than reflected. The height of the maximum tracks the reflected fraction at it, at about 0.6 of that fraction throughout. Larger probes intercept a wider range of angular momenta and leave less to reflect: at $\chi_p = 25$ and $T_i = T_e$, the maximum is 15.6, 7.2 and 1.5 percent at $R_p/\lambda_D = 0.5$, 1 and 2 and absent at 5. A maximum of the same kind appears in the orbital ion density that Filippov [9] computed for a micron-sized absorbing grain in an isothermal plasma.

We add one warning here. For attracted cold ions, the damped fixed-point iteration between the densities and Poisson's equation does not converge. The attracted density is a non-local function of the potential, the absorption radius flips between branches, and the iteration settles into a slowly growing oscillation that neither under-relaxation nor a diagonal Jacobian removes. Anderson mixing cures it, and we accept every point only if two radial meshes agree. At $T_i/T_e = 0.1$ two meshes disagree by a factor of five in the height of the maximum, and a single unconverged solve returns a smooth-looking profile with a maximum. The same non-uniqueness, the bistability of the absorption radius at $T_e/T_i \gg 1$ is the reason why only time-like relaxation, which the PAMAR species solvers perform, properly selects the physical solution [1].

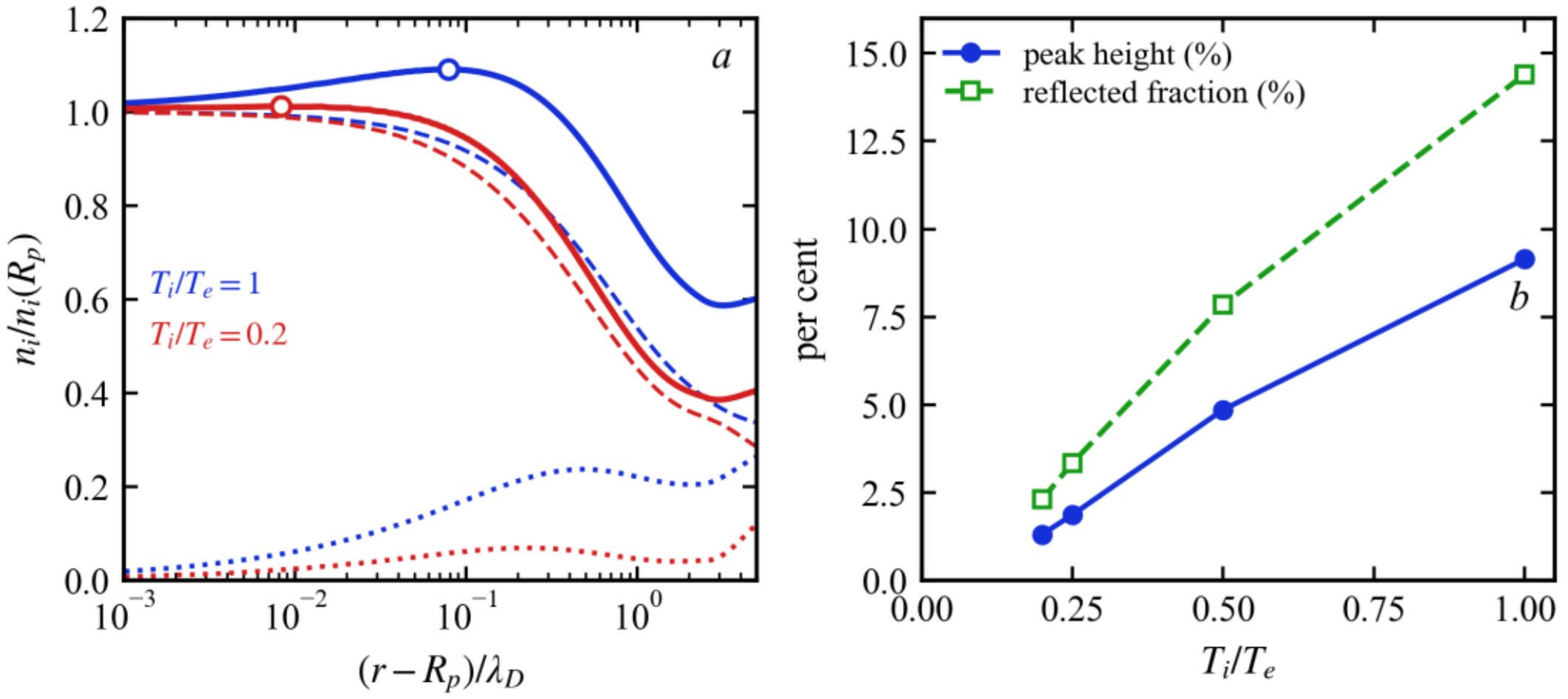

**Figure 5.** The maximum of the attracted-species density near the surface ($R_p = \lambda_D$, $\chi_p = 10$). (a) Total density (solid) with its inbound (dashed) and outbound (dotted) parts for $T_i/T_e = 1$ and 0.2, against the distance from the surface on a logarithmic scale; circles mark the maximum. (b) Height of the maximum and the reflected fraction at it against the temperature ratio.

## 6. Different solvers for electrons and ions

The choice of solver for each species follows from how its particles separate into groups, as shown by the comparison between the sphere and planar discharge in [1]. In both geometries, the potential divides the electrons by their total energy. The electrons below the wall barrier are trapped. They conduct no current and define the plasma density. Only the fast electrons above the barrier reach the surface and carry the current. Both groups must be computed accurately, because the density of the trapped group enters Poisson's equation and the current of the fast group must balance the ion current. The two geometries differ in the loss cone. At a planar wall, the cone of a fast electron is set by its energy alone and is the same at any distance from the wall. Around a sphere, it also narrows with the radius as $R_p^2/r^2$ and is much smaller. Inside the cone, the distribution is strongly anisotropic, and the 1d2v solution is needed. Outside the cone, the Fokker–Planck equation for $f_0(r,\varepsilon)$ is adequate and exact for the trapped group. This is why, with collisions, the electrons take the Fokker–Planck model and not the two-cone model at the angular-moment level (Section 3); without collisions, the two-cone model is exact for either species.

For the ions, the two-stream model is appropriate (Section 3): they move along the field at a planar wall and along orbits with angular momentum around a sphere, which the two-cone model follows (Section 6.4). Outside the few radii where angular momentum matters, the fluid model for the ions could work better around a sphere than in the capacitively coupled discharge of [1]: the field there is small, and ion heating between charge-exchange collisions is less significant than in the sheaths of that discharge.

### 6.1. Electrons without collisions: the graded mesh

The electron benchmark fixes the potential profile, so that any difference between the solvers comes from the electron kinetics. The potential energy of an electron is the Coulomb profile

$$U(r) = \chi_p T_e \frac{R_p/r - R_p/R_{out}}{1 - R_p/R_{out}}, \quad R_{out} = 10R_p, \tag{13}$$

with a Maxwellian of density $n_\infty$ entering at $R_{out}$; $\chi_p = 2$ repels the electrons, as for a dust grain or a probe below the plasma potential, and $\chi_p = -2$ attracts them. Table 2 lists the collected currents. The MoC agrees with the exact values to six digits. The repelled density falls to 0.068 $n_\infty$ at the surface, half the Boltzmann value because the outward hemisphere is empty there. The attracted density rises to about 1.33 $n_\infty$ near $1.5R_p$ and falls back to 0.96 at the surface, which is the mechanism of Section 5.4 on a larger scale.

**Table 2.** Collected electron current without collisions in units of the random current. The 1d2v values are the current through the wall face of the octree from the conservative face fluxes, with the exact boundary fluxes of Section 4.2; graded = logarithmic radial map, whose first cell at level 5 equals that of a uniform level-7 mesh.

| | **no field** | **repelled,** $\chi_p = 2$ | **attracted,** $\chi_p = -2$ |
|---|---|---|---|
| exact | 1 | 0.13534 | 2.9799 |
| MoC | 1.00000 | 0.13534 | 2.9799 |
| 1d2v, uniform, level 5 | 0.9514 | 0.0958 | 3.457 |
| 1d2v, uniform, level 6 | – | 0.1255 | – |
| 1d2v, uniform, level 7 | – | 0.1331 | – |
| 1d2v, graded, level 5 | 0.9683 | 0.1239 | 3.392 |
| 1d2v, graded, level 6 | – | 0.1328 | – |
| 1d2v, graded, level 7 | – | 0.1347 | – |

The 1d2v solver reproduces the profiles away from the surface and falls short at it. Its repelled current is 0.708 and 0.927 of exact on uniform meshes of levels 5 and 6 and 0.915, 0.981 and 0.995 on the graded meshes of levels 5, 6 and 7. A collisionless repelled species cannot deliver more than $e^{-\chi_p}$ whatever the shape of the potential. The corrected scheme respects that bound from below, and the deficit comes from the first-order extrapolation to the wall face analyzed in [1]. In contrast, the attracted current is 16 and 14 percent above exact on the uniform and graded level-5 meshes, and the graded wall cell barely moves it. The excess is not at the surface but in the bound region of Section 5.2, which the exact solution leaves empty, and numerical diffusion in velocity slowly fills it from passing orbits. The run drifts hundreds of transit times before it settles. We identified the surface error from the bound on the repelled current, checked while a run proceeds. Neither a profile comparison nor a convergence criterion on the potential would have found it.

The logarithmic map puts the cells where the loss cone opens: the wall cell is $0.075R_p$ against $0.28R_p$ on the uniform mesh of the same level. Beyond the wall cell, the density of the 1d2v solver and the temperatures along and across the radius follow the MoC. Grading has a price, since the cells beyond the layer are coarser than on the uniform mesh: the graded level-5 density over the Boltzmann factor is 0.035 below the MoC at $1.3R_p$, and level 6 brings the difference below 0.01. The collected current is the difference between an inward and an outward current that are each 450 times larger at $8R_p$, which is why the boundary fluxes must be exact. In cells, grading is worth one level of refinement: the deficit of the collected current on the graded meshes of levels 5 and 6, 8.5 and 1.9 percent, is close to that of the uniform meshes of levels 6 and 7, which have eight times as many cells (Table 2).

## 6.2. The loss cone and its filling by collisions

The distribution function explains why the wall cell dominates. In the cell next to the surface, the 1d2v solver spreads the edge of the loss cone over a third of the hemisphere, because orbits that graze the sphere and orbits that strike it differ by less than one cell there, and the scheme cannot separate them. One cell out the edge is sharp again. Refining the loss-cone edge improves the distribution there but not the collected flux; grading the radial axis toward the surface improves the flux.

Figure 6 shows the loss cone and its filling free of mesh effects, from the two exact solutions. Without collisions, a sharp boundary divides the plane of the velocity components along and across the radius: $f = f_\infty(\varepsilon)$ outside the loss cone and zero inside it (panels a to c, from the MoC). These panels show $f$ itself, which, outside the cone, is the Maxwellian of the total energy. The cone starts at the barrier speed, below which an electron cannot reach the surface, and narrows with the radius. For attracted electrons (panels m–o), the plane has a second empty region besides the shadow of the surface, the hole of the orbits that do not connect to the plasma. Its core is the disk of negative total energy, below $1.34v_T$ at $1.1R_p$, and since elastic collisions conserve energy, scattering does not fill it either; it is the electron counterpart of the ion hole of Section 5.3. The hole is nearly circular because the $1/r$ potential of the benchmark falls more slowly than the centrifugal term $\ell^2/r^2$, so the effective potential has no outer barrier. A screened potential has one, and the hole then widens across the radius: for $R_p/\lambda_D = 3$ and a probe potential of $5T_e$, it reaches $1.84v_T$ across the radius at $1.3R_p$ against $1.25v_T$ along it. With scattering (integration along the orbits) two things happen at once. The cone fills from its edge, and the distribution outside it falls below the Maxwellian, because the electrons that fill the cone are taken from the other orbits and are lost at the surface. At $r = 1.1R_p$ and $T_e$ above the barrier, $f/f_\infty$ on the axis of the cone rises from zero to 0.06, 0.12 and 0.17 at $\lambda = R_p$, $0.3R_p$ and $0.1R_p$, while the inbound value falls to 0.90, 0.70, and 0.46. At $3R_p$ and $\lambda = 0.1R_p$ the distribution is isotropic to 2 percent and lies 10 percent below the Maxwellian. The electrons below the barrier keep $f = f_\infty$ at any collisionality, since elastic collisions conserve their energy and none of their orbits ends on the surface. The depleted and nearly isotropic distribution above the barrier is what the Fokker–Planck model of Section 6.3 describes, and the layer where the cone is still visible is what it cannot describe.

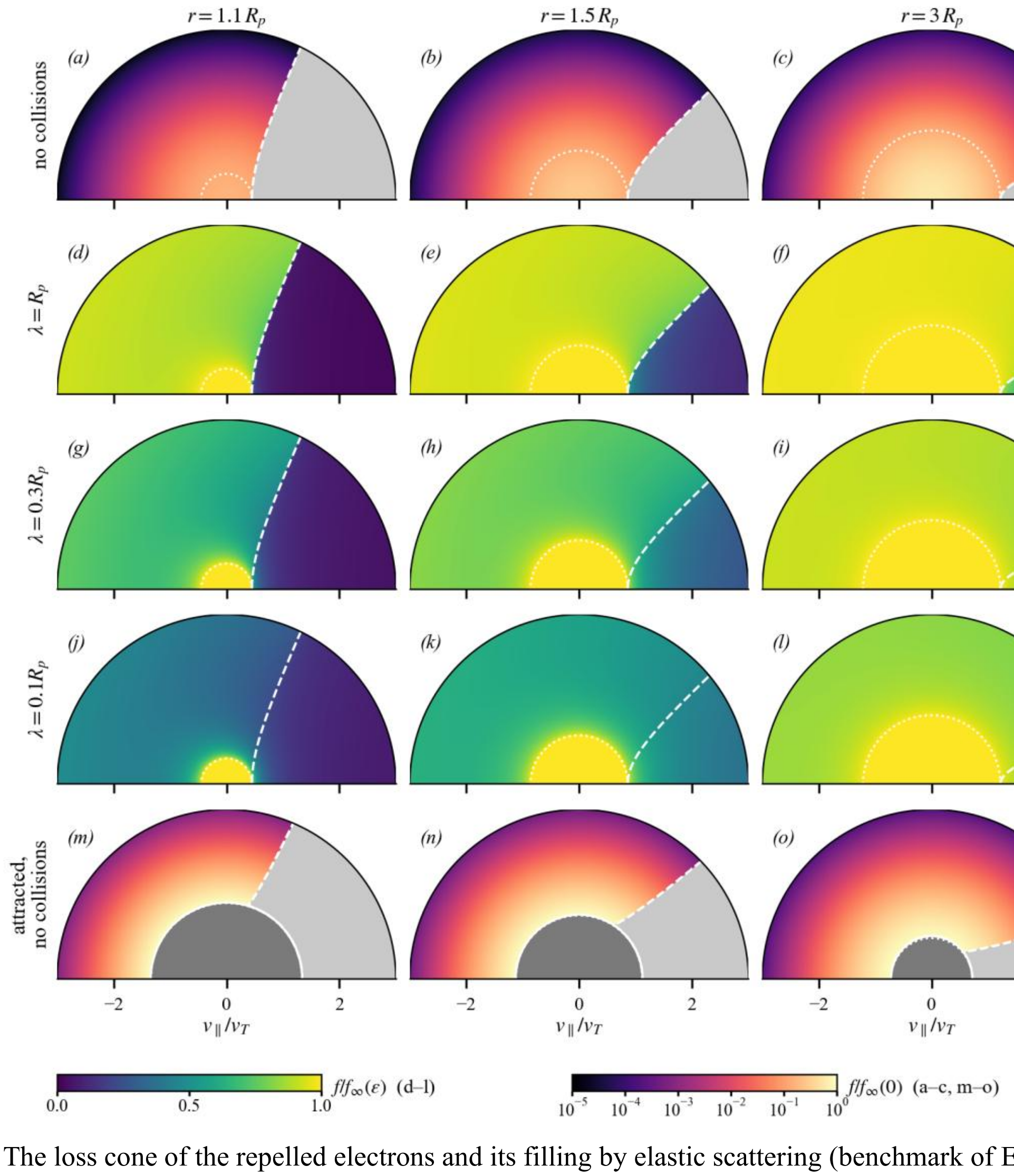

Figure 6. The loss cone of the repelled electrons and its filling by elastic scattering (benchmark of Eq. (13), $\chi_p = 2$). Distribution in the plane of the velocity components along (outward to the right) and across the radius, out to $3v_T$, at $r = 1.1$, 1.5 and $3R_p$: (a–c) without collisions, from the MoC, shown as $f/f_\infty(0) = e^{-\varepsilon/T_e}$ on a logarithmic scale with the empty orbits grey; (d–l) $\lambda = R_p$, $0.3R_p$ and $0.1R_p$, from the integration along the orbits, shown as $f/f_\infty(\varepsilon)$. Dashed: the collisionless edge of the loss cone; dotted: the barrier speed, below which the electrons cannot reach the surface. (m–o) Attracted electrons ($\chi_p = -2$) without collisions, from the MoC, on the scale of (a–c): dark grey with a solid edge, the hole of the orbits that do not connect to the plasma; light grey with a dashed edge, the absorption shadow; dotted, zero total energy.

## 6.3. Electrons with collisions: the Fokker–Planck model and the hybrid solver

With elastic scattering, the reduced electron model is the Fokker–Planck model of Section 3 [1], carried here to the absorbing surface. Figure 7 compares it with the integration along the orbits for the repelled electrons of the benchmark. The model is right in the diffusion limit and collects too much as the mean free path grows: 5 percent at $\lambda = 0.1R_p$, 18 percent at $0.3R_p$, 51 percent at $R_p$ and twice the exact current in free flight, since an isotropic distribution with a small gradient cannot represent the empty cone. The density profiles differ by less than the current (Figure 7(c)). At $\lambda = 0.3R_p$, it is within 0.03 of the exact density over the Boltzmann factor beyond $1.5R_p$, and the error is confined to the loss-cone layer.

The hybrid electron solver of [1] uses each description where it holds: the 1d2v octree in the loss-cone layer between the surface and an interface radius, and the Fokker–Planck model beyond it, coupled per velocity row through the partial currents, so that particles are conserved across the interface. On the graded level-6 mesh at $\lambda = 0.3R_p$ it gives the exact current within 0.2 percent with 56 percent fewer cells than the all-kinetic solver [1] (circle in Figure 7(a)). The interface is placed by the anisotropy seen in Figure 6, and Section 9.3 returns to it. For the electrons, the reduced model is thus the Fokker–Planck model outside the loss cone, with the 1d2v solver inside it.

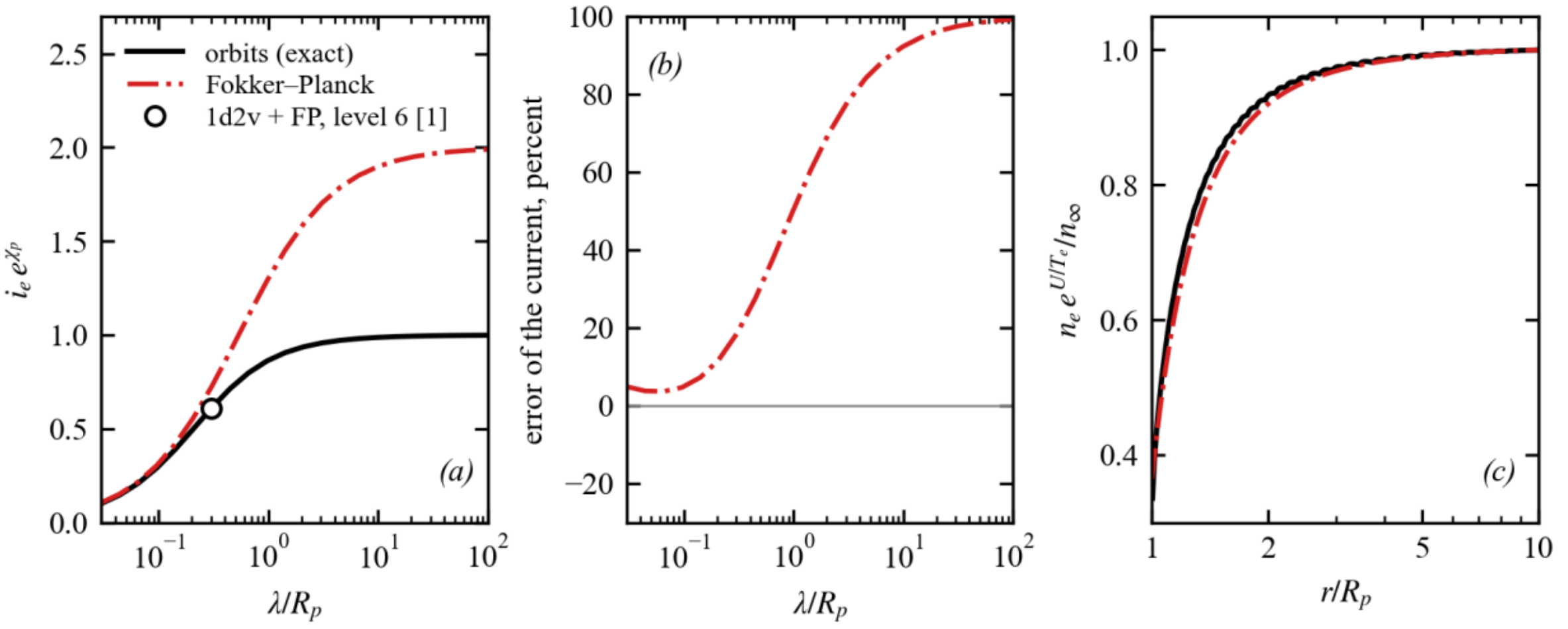

Figure 7. The Fokker–Planck model against the integration along the orbits (repelled electrons of the benchmark, $\chi_p = 2$, elastic scattering at constant $\nu_m$). (a) Collected current over the Boltzmann factor against the mean free path: orbits (solid), the Fokker–Planck model carried to the surface (dash-dotted); circle: the 1d2v + Fokker–Planck hybrid on the graded level-6 mesh [1]. (b) Error of the Fokker–Planck current. (c) Density over the Boltzmann factor at $\lambda = 0.3R_p$.

## 6.4. Ions: kinetic, reduced and fluid solvers

For the ions, the species solvers differ in what they carry, and we compare them with identical electrons, Poisson shells, boundary conditions, charge-exchange operator, and reservoir. The floating runs use artificial mass ratios (3 in this paragraph and in the Coulomb-regime state of Section 9.2, 100 in Table 3 and in the screened case of Section 9), and each case is identified by the floating potential $\chi_f$ that the MoC gives for it. The adaptive ion octree and the screened floating benchmark are reported in [1]: this case ($R_p/\lambda_{De} = 8, T_i = T_e$) floats at $\chi_f = 0.306$ against the MoC value 0.300. The Coulomb-regime case ($\lambda_{De} = 10R_p$, $T_e/T_i = 3$, $R_{out} = 10R_p$) floats at 0.339 and 0.384 on graded electron meshes of levels 5 and 6, compared with 0.363.

Table 3 compares the three ion solvers in the collisionless and collisional limits of the screened case with $\chi_f = 1.545$. The radial fluid over-collects by 50 to 69 percent, because with no angular momentum every ion that enters is collected, and floats 0.39–0.53 $T_e$ shallower than the kinetic solver; the collected current, not the floating potential, discriminates between ion models [1]. 1d1v(S) inverts the same barrier functions analytically rather than sampling them on a mesh, reproduces the collisionless anchor, and covers about 40 times more physical time than the kinetic solver in equal wall time. This makes it the candidate for the outer region of the hybrid in Section 9. Coupled to the same electron solver, it is within 0.6 percent of the MoC floating potential and 5 percent of the MoC current without collisions, and within 3 and 5 percent of the kinetic solver at $\lambda_i = 0.3R_p$ (Table 3). The latter values are averages over a persistent oscillation of the coupled run (period about 40 time units, amplitude 0.01 in the floating potential), which the kinetic run does not show. At fixed bias, the model follows the kinetic sweep of Figure 4. It collects 3.21, 2.50 and 1.74 $\Gamma_{th,i}$ at $\lambda_i/R_p = 10$, 3 and 1 against 3.17, 2.67 and 1.71, and its trapped bath holds 0, 3.5 and 38 percent of the ions in the domain. It does not reproduce the potential outside the sheath. The model attenuates the reservoir ions by charge exchange and does not follow the ions born on escaping orbits, so the outer part of the domain is left short of ions: $\chi$ at $R_{out}$ is 0.10, 0.15, and 0.26, versus 0.01, $-0.01$, and $-0.03$ in the kinetic runs. The two-cone model thus describes the orbital region within a few mean free paths of the sphere and requires a collisional model outside, which is the hybrid in Section 9.4.

**Table 3.** Floating potential and collected ion current (units of $\Gamma_{th,i}$) for the three ion solvers with the same 1d2v electron solver ($\chi_f = 1.545$ from the MoC without collisions, $T_i = T_e$, $R_p/\lambda_{De} = 8$, $R_{out} = 3R_p$ with the $r^{-2}$ tail). Kinetic and fluid values from [1]; the 1d1v(S) values at $\lambda_i = 0.3R_p$ are averages over the oscillation described in Section 6.4.

| **ion solver** | $\chi_f, \lambda_i \to \infty$ | $i_i, \lambda_i \to \infty$ | $\chi_f, \lambda_i = 0.3R_p$ | $i_i, \lambda_i = 0.3R_p$ |
|---|---|---|---|---|
| MoC (exact) | 1.545 | 2.134 | – | – |
| 1d2v kinetic | 1.504 | 2.09 | 2.671 | 0.66 |
| 1d1v(S) | 1.554 | 2.03 | 2.59 | 0.69 |
| fluid | 1.114 | 3.13 | 2.142 | 1.11 |

# 7. The current–voltage characteristic of the collisionless probe

## 7.1. Both branches at a prescribed drop

Here, we prescribe the drop and compute both currents for that drop with the potential solved self-consistently. The characteristic is then assembled point by point, with the ion-collecting branch at $\chi_p > 0$ and the electron-collecting branch at $\chi_p < 0$ (Figure 1(b)). Three exact statements appear in the output. First, the repelled branch is $e^{-\chi_p}$ to $2 \times 10^{-4}$ over $\chi_p = 0.5$ to 10, since for a monotone barrier Liouville's theorem saturates the Boltzmann relation, whatever the shape of the potential. Second, at equal temperatures the two branches are mirror images. Third, in a Coulomb potential the attracted current tends to the OML value $1 + |\chi_p|$ as the outer boundary moves away, short of it by 0.7 percent at $R_{out} = 10R_p$, with the deficit falling as $(R_p/R_{out})^2$ (Figure 1(c)). A calculation that quotes an OML current should therefore quote its outer boundary as well.

The first statement underpins the Druyvesteyn formula. For an isotropic distribution and a monotone retarding potential, the inbound orbits reach the surface with their reservoir value, so the repelled current per unit area is

$$j_e(V) = \frac{2\pi e}{m^2}\int_{eV}^{\infty}(\varepsilon - eV)f(\varepsilon)\,d\varepsilon, \quad \frac{d^2 j_e}{dV^2} = \frac{2\pi e^3}{m^2}f(eV), \tag{14}$$

with $V$ the retarding potential and $\varepsilon$ measured from the plasma potential [10]. Nothing in Eq. (14) depends on the sheath thickness, the probe size or the shape of the potential, which is why the collisionless characteristic carries the EEDF exactly. Collisions inside the region from which the current is drawn break it. Section 8 measures their effect on the current; recovering the distribution from a collisional characteristic is left to a separate paper.

## 7.2. The electron-collecting branch and the ion temperature

On the electron-collecting branch, electrons are attracted and ions repelled by a barrier $|\chi_p|T_e/T_i$. At $R_p/\lambda_D = 1$ and $\chi_p = -25$ the collected electron current is 18.11 $\Gamma_{th,e}$ for $T_e/T_i = 10$, 69.7 percent of the OML ceiling of 26, and 21.91 for $T_e/T_i = 1$. The repelled-ion temperature moves the current by about 20 percent at the deepest bias, through the ion space charge that shapes the electron sheath, so that a right-branch current quoted without $T_e/T_i$ is ambiguous. The sheath extends to about $5R_p$, where $|\chi|$ falls to 0.1, and the ions are excluded ($n_i < n_\infty/2$) out to $5.3R_p$. Holding $\chi_p = -25$ and raising $\lambda_D/R_p$ through 0.3, 1, 3, 10, 30 and 100 gives 0.363, 0.697, 0.910, 0.987, 0.998 and 0.9997 of the OML value. Thus, OML is the $\lambda_D \gg R_p$ corner of the same calculation, and $\lambda_D \sim R_p$ is the crossover where neither OML nor the thin-sheath limit is accurate. As on the other branch, trapped orbits are assumed empty; here they hold electrons, and where the solution shows a well, the current is an upper bound.

## 7.3. The ion-collecting branch with cold ions

For cold ions, Laframboise tabulates the current in units built with the electron temperature, which differ from the ion thermal units used here by $(T_e/T_i)^{1/2}$, a factor of 3.16 for $T_i/T_e = 0.1$ that otherwise looks like a disagreement in physics. In consistent units, the MoC reproduces all eleven entries of Table 5b block ($T_i/T_e = 0.1$, $R_p/\lambda_D = 10$, $\chi_p = 0.1$–$10$) within 1 percent. The cold-ion branch needs the Anderson-mixed iteration of Section 5.4 and a larger domain than the repelled branch: at $R_{out} = 20R_p$ the residual stalls near $10^{-2}$, a domain-truncation effect, while $R_{out} \geq 40R_p$ converges below $10^{-4}$. The profiles behind these currents show that an attracted species can be depleted at the wall. At $\chi_p = 5$ the ion density is 0.413 $n_\infty$ at the surface, because acceleration increases their speed faster than geometric convergence increases their flux, and the outgoing half of their distribution is missing.

## 7.4. Screening and the floating potential

At fixed bias, the screening length controls the answer. At $\chi_p = 5$ and $T_i = T_e$, as $\lambda_D/R_p$ runs from 0.2 to 20, the logarithmic slope of the potential at $2R_p$ goes from $-3.65$ (a thin sheath falling far faster than the Coulomb law) to $-1.12$. The radius at which $\chi$ falls to unity moves out from 1.5 to $4.1R_p$, and the collected ion current rises from 3.96 to 6.00 $\Gamma_{th,i}$, the OML ceiling $1+\chi_p$.

The mass ratio enters only through Eq. (10). We compute the characteristic once, and the floating point slides along it. At $R_p = \lambda_D$ and $T_i = T_e$ it sits at $\chi_f = 0.293$, 1.427, 2.523 and 4.028 for $M/m = 3$, 100, 1836 ($H^+$) and $7.3 \times 10^4$ ($Ar^+$) (the triangles in Figure 1(b) mark the first, third and fourth). At $R_p/\lambda_{De} = 8$ the floating potential is 0.300 for $M/m = 3$ and 1.545 for $M/m = 100$, independent of where the outer boundary is put once the $r^{-2}$ tail holds it (1.545 at $R_{out} = 3$ and $12R_p$). This justifies the artificial mass ratios of the kinetic runs, since at a fixed bias their profiles are the physical ones, and the floating state at the real mass ratio is recovered by holding the bias at the value the balance predicts.

## 7.5. The characteristic from the coupled solvers

Sections 7.1–7.4 computed the characteristic with the MoC alone. Here the coupled solvers compute it, with the bias held at twelve values from $\chi_p = -5$ to 5 and the potential, both densities and both currents found by the time-dependent solution itself: the 1d2v solver for both species on the graded level-5 meshes of Table 2, and the two-cone model 1d1v(S) for the ions with the same electrons. The case is the screened one of Section 6.4 ($R_p/\lambda_{De} = 8$, $T_i = T_e$, $R_{out} = 4.5R_p$ with the $r^{-2}$ tail). Each run starts from the MoC state at its bias and stops when both wall currents are steady, after 11–26 minutes on one core for 1d2v and 5–12 minutes for 1d1v(S). Figure 8 shows the result.

The 1d2v characteristic follows the MoC on both branches. The attracted current is 4.4 and 4.1 percent low at $|\chi_p| = 0.5$ and 1, 1.6 percent low at 2, and within 1.4 percent at 3–5, with the two branches mirror images to 0.5 percent, as the exact solution requires at $T_i = T_e$ (Section 7.1). The repelled current is 7–13 percent low at every bias. That is the wall-cell deficit of the level-5 mesh measured at a prescribed drop in Table 2 (0.1239 against 0.1353 at $\chi_p = 2$), so the self-consistent field adds nothing to the error of the species solvers; by Table 2, level 7 reduces it to 0.5 percent.

Figure 9 compares the profiles of one run of the series, $\chi_p = 2$, with the MoC. The potential and both densities lie on the MoC curves. So do the moments that the densities do not test: the drift velocity and the temperatures along and across the radius, computed from the two octrees, agree with the moments of the MoC distribution within 0.03 of the species temperature at every radius, and the ion drift within 1 percent. The repelled electrons keep $T_\perp = T_e$ and lose parallel temperature toward the wall, down to $0.5T_e$ at the surface, where the outbound hemisphere is empty. The attracted ions arrive with a drift of 1.6 $(T_i/M)^{1/2}$, $T_\parallel = 0.4T_i$ and $T_\perp = 1.9T_i$, since the fall narrows the radial velocities and the conservation of angular momentum raises the transverse ones. The ion anisotropy persists beyond $3R_p$ ($T_\parallel/T_\perp = 0.86$ at $3.3R_p$), while the electrons are isotropic to 5 percent beyond $1.8R_p$. This difference is what the sensors of Section 9 measure.

Coupled to the same electrons, the two-cone model gives the repelled-ion branch within 2 percent down to 0.007 at $\chi_p = -5$, where the octree is 13 percent low, because it inverts the barrier analytically and no wall cell enters. On the ion-collecting branch, it is 3.0 percent low at $\chi_p = 0.5$ and 4.9 percent low at 5, an almost uniform offset that matches the 5 percent deficit of the floating case of Table 3. The electron current computed in the field of the model's ions stays within 5 percent of the MoC on the electron-collecting branch, 2–4 percent above it at $|\chi_p| \geq 3$ where the model's sheath is slightly thinner. At fixed bias, the relaxation time of the moments that 1d1v(S) passes to the field solver must be short.

The dashed curves of Figure 8(a) give the opposite anchor, the drift–diffusion limit of the same characteristic: the steady continuum equations for both species with Poisson's equation at the same screening, extrapolated to an infinite domain, in units of each species' own field-free flux. Collisions flatten both ends. The attracted flux grows 2.3 times between $\chi_p = 0$ and 5 against 3.4 times without collisions, since the ions diffuse toward the sphere instead of falling onto it, and the repelled flux at $|\chi_p| = 5$ is 0.072 of its field-free value against $e^{-5} = 0.0067$, since the barrier is crossed by diffusion rather than by energy, so that the Druyvesteyn exponential (Section 7.1) is lost. Section 8 follows the passage between the two anchors with the mean free path.

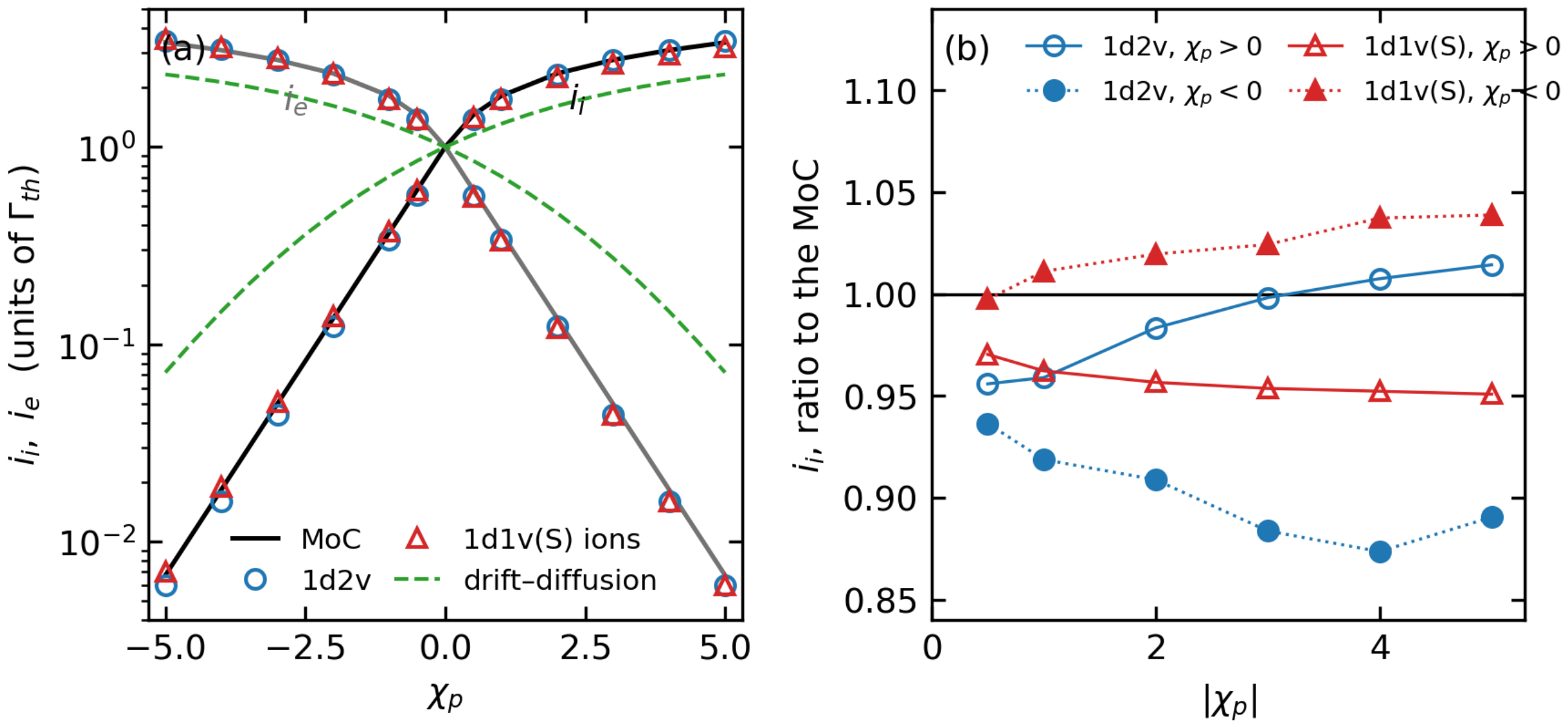

Figure 8. The screened sphere ($R_p/\lambda_{De} = 8$, $T_i = T_e$) from the coupled solvers at twelve fixed biases. (a) Collected ion and electron currents, in units of the thermal flux of each species: 1d2v for both species (circles), 1d1v(S) ions with 1d2v electrons (triangles), MoC (lines); dashed, the drift–diffusion limit in units of its own field-free flux. (b) The ion current as a ratio to the MoC, attracted ($\chi_p > 0$) and repelled ($\chi_p < 0$).

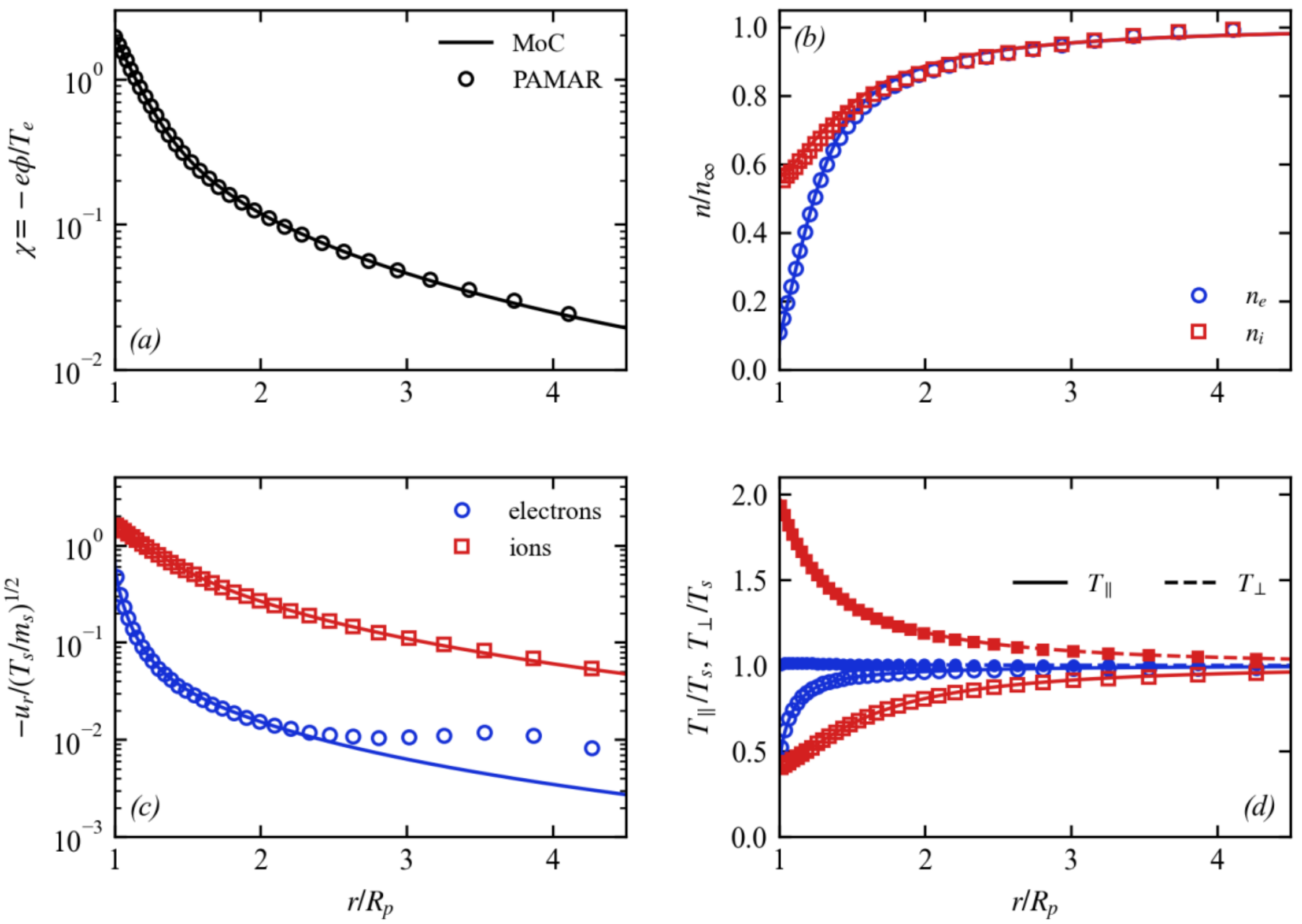

Figure 9. Profiles at the fixed bias $\chi_p = 2$ ($R_p/\lambda_{De} = 8, T_i = T_e, R_{out} = 4.5R_p$): 1d2v for both species (symbols) and the MoC (lines); blue, electrons; red, ions. (a) Potential. (b) Densities. (c) Inward drift velocity in units of $(T_s/m_s)^{1/2}$; the electron drift beyond $2.3R_p$ is below the resolution of the level-5 velocity mesh. (d) Temperatures along (open symbols, solid lines) and across (filled symbols, dashed lines) the radius, in units of the species temperature.

## 7.6. The far field and ion collisions

Near the sphere, the potential is screened and falls faster than the Coulomb law (Section 7.4). Far from it, the potential follows the Coulomb law in neither limit. Without collisions, the absorbing sphere removes the ions of the outbound loss cone, whose solid angle falls as $R_p^2/r^2$, and the potential that keeps the plasma neutral falls as $r^{-2}$ [4,6]. With ion–neutral collisions, the ions reach the sphere by drift and diffusion, and the potential beyond the mean free path is Coulomb-like, $r^{-1}$, with an effective charge set by the collected flux rather than the sphere's charge [23].

Neither tail is within reach of the coupled time-dependent runs. On $R_{out} = 4.5R_p$, the potential is still in the screening region, and on a domain of tens of Debye lengths, a charge imbalance of 1 percent, the accuracy of the densities on a velocity mesh, produces a potential of order $T_e$, larger than the tail itself. We therefore computed the far field with two stationary solvers that have no velocity mesh. Without collisions, the two-cone model evaluates both densities by the same quadrature, so that their errors cancel where the plasma is neutral, and Poisson's equation is solved on geometric radial nodes. With charge exchange, the two-cone model is not adequate far from the sphere (Section 6.4). There, we integrate the ion kinetic equation along the exact orbits on a lattice of energy and angular momentum, bound orbits included, with the birth rate $S_{CX}(r)$ found self-consistently, so that the plasma far from the sphere stays Maxwellian at any number of mean free paths.

Figure 10 shows the result for a small sphere ($R_p = \lambda_D$, $\chi_p = 3$, $T_i = T_e$, $R_{out} = 200R_p$). Without collisions, the logarithmic slope of the potential is $-2.0$ between 10 and $30\lambda_D$; the potential agrees with the MoC within 1 percent and does not change with the exponent of the outer condition or with $R_{out}$. At $\lambda_i = 10R_p$, the potential follows the collisionless curve inside the mean free path and leaves it beyond: the slope is $-1.2$ between 10 and $40\lambda_D$, and at $40\lambda_D$ the potential is eleven times the collisionless one. At $\lambda_i = 30R_p$, the departure begins farther out. Colder ions ($T_e/T_i = 3$) give the same slopes. The exponent of the outer condition does not matter here either, since the plasma is neutral at $R_{out}$; the reservoir pins the potential there, which bends the collisional curves down within a factor of two of $R_{out}$. Khrapak et al. [23] obtained the two laws for a grain much smaller than the Debye length; the present case shows them for a sphere of the size of the Debye length.

We checked both tails against Filippov's kinetic theory [9]. For his micron-sized grain at the floating potential in an isothermal plasma ($T_i = T_e$, $R_p/\lambda_{De} = 0.026$), the collisionless solver gives the coefficient of the $r^{-2}$ tail within 0.3 percent of his values for the five noble gases. The coefficient decreases as the sphere grows: at $\chi_p = 3$, the product $\chi r^2/R_p^2$ is 0.88 at $R_p/\lambda_D = 0.1$ and 0.75 at $R_p = \lambda_D$, because the slow ions, which dominate the tail, no longer have the orbital cross section. With collisions at a constant frequency, the collision model of [9], the Coulomb part of the potential is proportional to $1/\lambda_i$ for $\lambda_i$ between 30 and $300R_p$. Beyond the mean free path, its amplitude is within 7 percent of the drift–diffusion value set by the collected flux [9]. Inside the mean free path, it is 0.6 of that value.

The same solvers give the collected current through the transition from orbital to radial ion motion. Without collisions, the ion current at $\chi_p = 3$ falls from the OML value $1 + \chi_p T_e/T_i$ as the sphere grows: 4.00, 3.99, 3.87, 3.41 and 2.61 $\Gamma_{th,i}$ at $R_p/\lambda_D = 0.1$, 0.3, 1, 3 and 10 for $T_i = T_e$, and 9.98, 9.91, 9.07, 6.75 and 4.38 for $T_e/T_i = 3$. The radial-motion limit [5] assumes ions without angular momentum. Cooling the ions does not produce it in a collisionless plasma, where a small sphere keeps the orbital current at any ion temperature. Charge exchange does, since it replaces an orbiting ion by a slow one that falls toward the sphere along the radius [7], and the passage between the radial and orbital theories has been observed with cylindrical probes in weakly collisional plasmas [24]. At $R_p = \lambda_D$, the current stays within 5 percent of its collisionless value down to $\lambda_i = 10R_p$ and falls to 3.02 and 7.09 $\Gamma_{th,i}$ at $\lambda_i = 3R_p$ for $T_e/T_i = 1$ and 3, where ion mobility limits the collection. At these temperature ratios, the enhancement of the current by rare collisions, which is large for the cold ions of dusty plasmas [7,25], does not exceed the accuracy of the orbit lattice, a few percent. For the screened sphere of Figure 4, the current falls 1.96 times between the collisionless limit and $\lambda_i = R_p$.

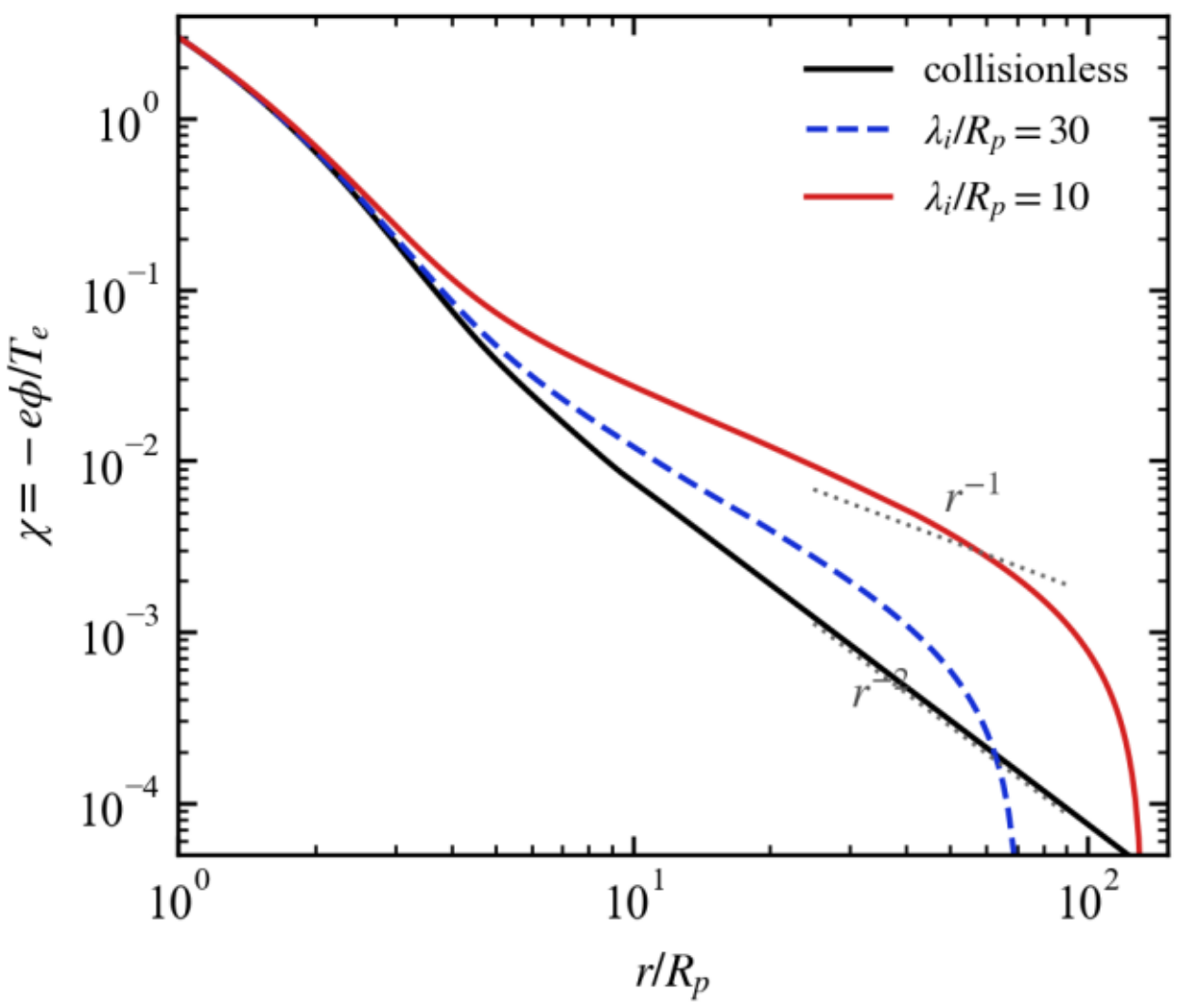

Figure 10. The potential far from a small sphere ($R_p = \lambda_D$, $\chi_p = 3$, $T_i = T_e$, $R_{out} = 200R_p$): collisionless (black) and with charge exchange at $\lambda_i = 30R_p$ (dashed) and $10R_p$ (red), from the stationary solvers of Section 7.6. Dotted: the laws $r^{-1}$ and $r^{-2}$.

## 8. Electrons with elastic scattering

With elastic scattering, the classification of Section 5 no longer applies, and the reference is the integration along the orbits, Eq. (9). Figure 11 follows the current of the repelled electrons of the benchmark, Eq. (13) with $\chi_p = 2$, from free flight to diffusion. Scattering reduces the current, since an electron entering with enough energy is turned away more often than an electron in the reflected band is turned into the loss cone. At $\lambda \ll R_p$ the current approaches the drift–diffusion value with $D = T_e/m\nu_m$, proportional to $\lambda$. The reduction is 13 percent at $\lambda = R_p$ and 39 percent at $\lambda = 0.3R_p$.

The 1d2v solver on the graded level-5 mesh carries its first-cell deficit, 0.915 of the collisionless current, at every mean free path. Divided by its own collisionless value, its current agrees with the reference to 1 percent at every mean free path down to $0.3R_p$, although there the first cell, $0.075R_p$, is a quarter of the mean free path. The wall-cell error scales the current but not its dependence on collisions, so that a mesh too coarse for the absolute current can still be trusted for the trend. Scattering fills the loss cone and depletes the other regions most strongly near their edges (Figure 11(c)). The density is much less sensitive than the current (Figure 11(b)). Because elastic scattering conserves energy, every energy group is a separate linear problem, and the same orbit solution gives the current collected from any isotropic distribution. This is the kernel of collisional probe diagnostics [11,12,14], which we treat in a separate paper.

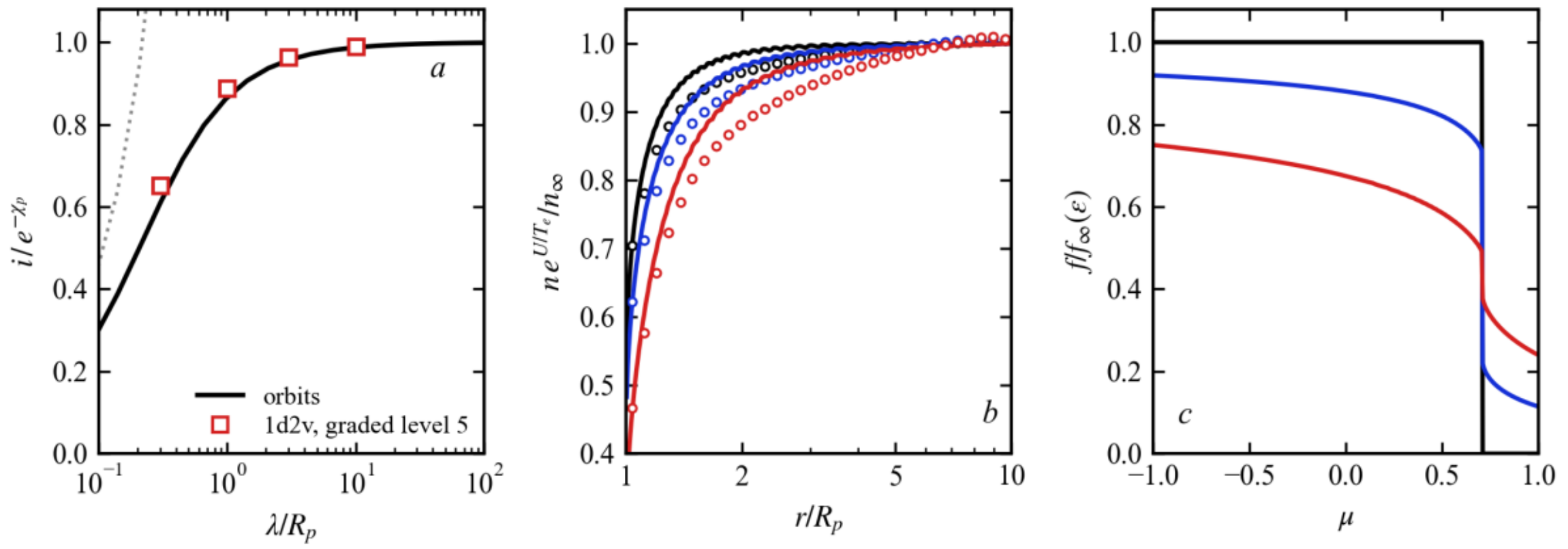

**Figure 11.** Repelled electrons with elastic scattering. (a) Collected current versus mean free path: along the orbits (black), 1d2v on the graded level-5 mesh divided by its collisionless value (symbols); dotted: the drift–diffusion limit. (b) Density divided by the Boltzmann factor for no collisions (black), $\lambda = R_p$ (blue) and $0.3R_p$ (red): orbits (solid), 1d2v (circles). (c) Angular distribution at $r = 1.2R_p$ at $T_e$ above the barrier, from the integration along the orbits.

# 9. Algorithm refinement cell by cell

## 9.1. From the species to the cell

In the companion paper, the model of a species is chosen at build time for the whole domain, except in two demonstrations of cell-by-cell selection [1]. The comparisons in Sections 5–8 say where each model is adequate, and the answer is local. Orbital focusing, the loss cone, and the trapped island are confined to a few probe radii; far from the body, both species are close to drifting Maxwellians. Our Unified Flow Solver selected Boltzmann and continuum solvers cell by cell from continuum-breakdown criteria for neutral gases [16,17]. For the plasma, the selection must be made per species because electrons and ions break down in different places. The PAMAR interface makes that possible without touching the field solver, since a hybrid solver contains a kinetic and a reduced solver and returns to the field solver, shell by shell, the moments of whichever owns the shell [1]. The sphere is a good demonstration because exact references exist for both species and because the sensors can be evaluated on an exact state before hybrid runs.

We consider two demonstrations of cell-by-cell selection on the sphere. The first concerns the electrons and is briefly introduced in [1]. The 1d2v solver is kept in the loss-cone layer, and a Fokker–Planck model of the isotropic part serves elsewhere (Section 9.3). The second demonstration concerns the ions and couples a fluid model, where the ions are a drifting Maxwellian, to a kinetic solver, where orbital motion and trapping define the collected current. The runs use the 1d2v octree, and Section 9.4 outlines the two-cone 1d1v(S) model as its replacement. The fluid–kinetic hybrid of the planar discharge is the template for the second; the first realizes the angular-moment level of the electron hierarchy of [1].

## 9.2. Sensors

Four sensors cover the physics of Sections 5–8. Each is a number per field shell and species, and a cell is owned by the kinetic solver when any of them exceeds its threshold.

- Anisotropy, $\delta_A = \left|T_\parallel/T_\perp - 1\right|$, from the second moments. It measures what the fluid closure throws away: the loss cone, which removes outgoing particles, and orbital focusing, which converts radial into transverse energy. It applies to both species.
- Drift, $M_i = |u_i|/c_s$ with $c_s = (T_e/M)^{1/2}$, the Bohm criterion of the planar hybrid; it marks the edge of a collisionless sheath, where the fluid model meets its sonic-point singularity.
- Trapping, the fraction of the $(v, \mu)$ plane occupied by bound orbits, from the barrier functions (11) of the current potential. Where it is nonzero and $\lambda_i$ is finite, charge exchange fills orbits that only the kinetic solver or the trapped bath of 1d1v(S) can carry. Weighted with the Maxwellian of the gas, it is the fraction of the ions born by charge exchange at a given radius that are born on bound orbits: in the Coulomb-regime floating state it peaks at 6.5 percent at $1.5R_p$, whereas in the screened state it is below 0.1 percent everywhere; there the island of the charge-exchange runs is dug by the space charge of the trapped ions themselves, so the sensor has to be re-evaluated on the evolving potential rather than on the collisionless start.
- Knudsen layer, $(r - R_p)/\lambda_s < k$: within a few mean free paths of an absorbing wall the distribution is anisotropic however collisional the plasma is (Section 8), and outside it the angular-moment (diffusion) level is exact to order $\lambda_s/L$.

The first three can be computed by the field solver from the potential alone through the MoC, making them predictors available in every shell, including those owned by the reduced model, where the kinetic moments are not. Inside the kinetic region, they are corrected from the kinetic moments. Figure 12 shows the first two sensors evaluated on the exact collisionless floating states of the two benchmark cases. One can see that in the screened case ($R_p/\lambda_{De} = 8$, $T_i = T_e$, $\chi_f = 1.545$, $R_{out} = 12R_p$) the ion drift exceeds $c_s$ only within $1.15R_p$ of the center, i.e. in the sheath, but the ion anisotropy stays above 0.1 out to $3.6R_p$ and above 0.05 out to $5.2R_p$, while the electron anisotropy falls below 0.05 at $2.1R_p$. The two species therefore require kinetic regions that differ by a factor of 2.5 in radius. The Bohm criterion that serves the planar discharge would place the ion interface inside the region where the fluid ions over-collect by 50 percent (Section 6.4), so that around a sphere the anisotropy, not the drift, is the appropriate sensor. In the Coulomb-regime case ($\lambda_{De} = 10R_p$, $T_e/T_i = 3$, $\chi_f = 0.363$) the ions never reach $c_s$ outside the first cell, and the anisotropy thresholds of 0.05 fall at $4.9R_p$ for the ions and $3.5R_p$ for the electrons.

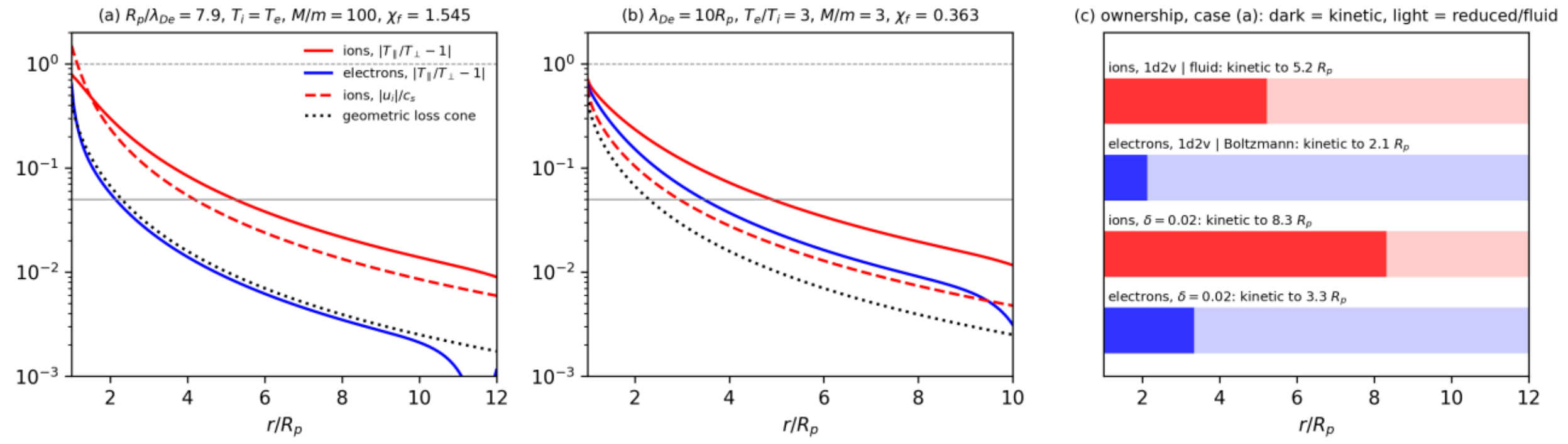

**Figure 12.** Breakdown sensors on the exact collisionless floating states (MoC, trapped orbits empty). (a) Screened case, $R_{out} = 12R_p$ with the $r^{-2}$ tail: pressure anisotropy of the ions (red) and electrons (blue), ion drift over the Bohm speed (red dashed) and the geometric loss-cone solid angle $(1 - \sqrt{1 - R_p^2/r^2})/2$ (dotted); grey lines:

thresholds 0.05 and 1. (b) Coulomb-regime case, $R_{out} = 10R_p$. (c) Ownership implied for case (a) by an anisotropy threshold of 0.05 and 0.02: kinetic (dark) and reduced (light) shells for each species.

## 9.3. Electrons, kinetic in the loss-cone layer

The electron demonstration is reported in [1]; we summarize it here because it was designed on the orbit solution of Section 8. Below the wall barrier $\varepsilon_w = \chi_p T_e$ the electrons are trapped by the potential, and since elastic collisions conserve their energy, their steady distribution is isotropic $f_\infty(\varepsilon)$ at any collisionality (Figure 13(a), below the red line). Above it, the loss cone keeps the distribution anisotropic within a layer whose thickness the sensor $\delta_A$ sets and which thickens with the energy (Figure 13(a), white contours). That the distribution next to an absorbing wall is depleted above the wall energy and unchanged below it is known from nonlocal theory [26]; the orbit solution gives the radial extent of the depleted layer. The hybrid electron solver keeps the 1d2v octree out to the interface and beyond it solves the Fokker–Planck (Lorentz) model of the isotropic part $f_0(r, \varepsilon)$, steady per energy group. The two are coupled per velocity row through the outward partial current of the octree and the half-range inflow of the P1 state, which conserves particles exactly across the interface. The division is thus made twice: in energy at the wall barrier, between the trapped electrons, known analytically, and the escaping ones, and in radius at the interface, which spans all escaping energies. The loss cone itself lies inside the layer and is refilled there by scattering; the sensor finds the radius beyond which the refilling is complete.

Two findings of the runs correct the design. First, the trapped region cannot be masked once the electrons collide, since the surface of the wall barrier cuts the octree cells in a staircase and neither a reflecting nor a zero-flux edge is conservative there; every velocity row at the interface, trapped rows included, is therefore coupled to the Fokker–Planck model. Second, the hybrid converges with the mesh as the all-kinetic solver does: 2–7 percent above it on the graded level-5 mesh, 0.7 percent on level 6, where its current is within 0.2 percent of the exact one. With the self-consistent potential, the result is the same, with 56 percent fewer electron cells and 11 percent less total time [1].

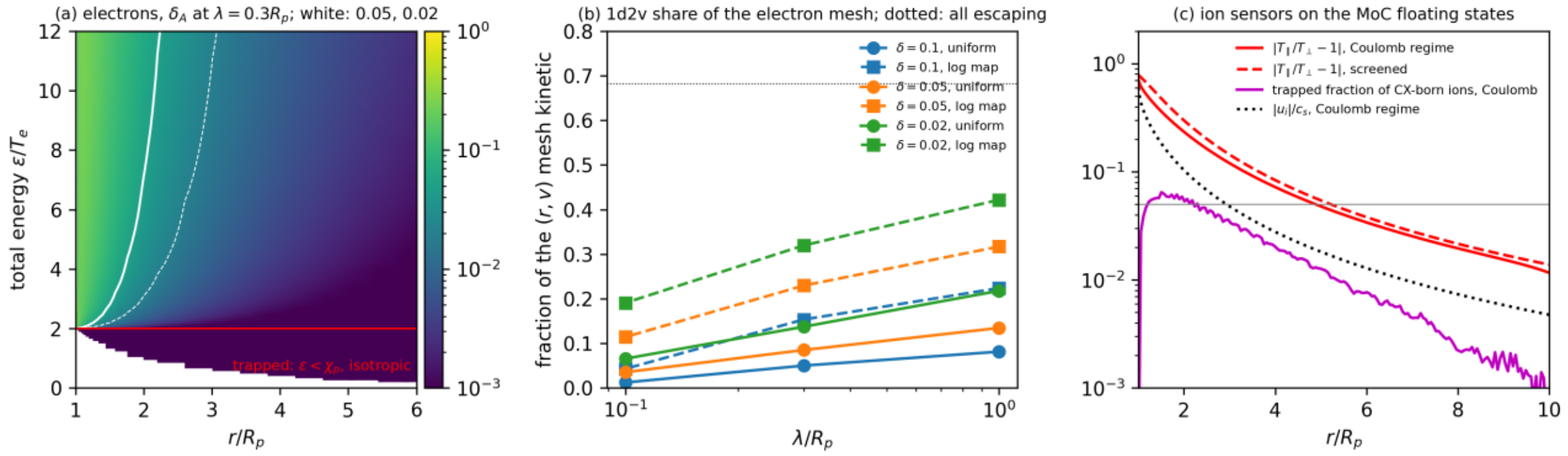

**Figure 13.** The two demonstrations on the exact solutions. (a) Electrons: the anisotropy $\delta_A = |T_\parallel/T_\perp - 1|$ of the orbit solution in the $(r, \varepsilon)$ plane for the benchmark of Section 6 at $\lambda = 0.3R_p$; red line: the wall barrier $\varepsilon_w = \chi_p T_e$, below which the electrons are trapped and exactly isotropic; white: the contours 0.05 (solid) and 0.02 (dashed) that bound the kinetic layer. (b) The share of the $(r, v)$ mesh owned by the 1d2v solver against the mean free path for three thresholds, on a uniform (circles) and a logarithmically graded (squares) radial mesh; dotted: all escaping electrons. (c) Ions: pressure anisotropy on the MoC floating states of the Coulomb-regime (solid) and screened (dashed) cases, the trapping sensor (fraction of charge-exchange-born ions on bound orbits, Coulomb regime) and the ion drift over the Bohm speed; grey line: 0.05.

## 9.4. Ions, fluid outside and kinetic inside

The ion demonstration is the spherical version of the planar hybrid, with three changes imposed by the sphere. First, the sensor is the pressure anisotropy, not the drift. Figure 13(c) and Figure 12 show that the drift exceeds the Bohm speed only inside the sheath, within $1.15R_p$ in the screened case and nowhere in the Coulomb regime, while the anisotropy stays above 0.05 out to 5.2 and $4.9R_p$ and the trapping sensor peaks at $1.5R_p$. An interface placed by the Bohm criterion would sit inside the region where the fluid ions over-collect by 50 percent (Section 6.4). Second, the octree beyond the interface is masked rather than coarsened, since coarsening saved no run time in the planar hybrid [1]. Third, the state the fluid hands to the kinetic region is a drifting Maxwellian of the fluid $n$, $u$ and $T_i$, which misrepresents the inbound hemisphere by the loss-cone depletion of the outbound one, a relative error of order $\delta_A$ at the interface. The threshold is therefore also the accuracy of the inflow, which is why the runs scan it. The moment-matching inflow of the planar hybrid [1] is the alternative; the runs below do not use it.

**Fluid–1d2v.** The hybrid solver combines ion solvers with sensors computed from its own moments; the kinetic moments $T_\parallel$ and $T_\perp$ are passed over the leaves. The first case is the collisionless floating sphere of Figure 12(a) ($R_p/\lambda_{De} = 8$, $T_i = T_e$, $R_{out} = 12R_p$, $\chi_f = 1.545$ from the MoC), for which the all-kinetic run and the MoC are the references. We run the hybrid at the anisotropy thresholds $\delta = 0.1$, 0.05 and 0.02, with the interface starting at 3.6, 5.2 and $8.3R_p$. The measures are the floating potential, the collected ion current, the potential across the interface (a kink is a coupling error, since the exact solution is smooth) and the kinetic cell count. The second case is the floating charge-exchange series of [1], whose ends are in Table 3 ($\lambda_i/R_p = 3$, 1, 0.3). There, the fluid region improves as collisions isotropize the ions, and ownership is re-evaluated during the run, testing the dynamic selection. The trapping sensor of Section 9.2 is not implemented in these runs.

In a first series, the ions advanced on their own clock (Section 4.3), and after about two ion transits of $R_{out}$ the floating potential began to oscillate and drift; a lock-step restart from the same state returned to $\chi_f = 1.57$ within 100 time units. We therefore reran the series with lock-step clocks to 2500 time units, about fifteen ion transits of $R_{out}$ (Figure 14). The interface sits at 3.3, 4.3 and $6.7R_p$ for $\delta = 0.1$, 0.05 and 0.02, 0.3–$1.6R_p$ inside its position on the MoC state (3.6, 5.2 and $8.3R_p$), and does not move during the runs. Between 500 and 1000 time units, the runs are quasi-steady. The all-kinetic run (run A) holds $\chi_f = 1.553 \pm 0.003$, 0.5 percent above the MoC value 1.545, with the ion current 2.04 (4 percent below the MoC value 2.134, the surface-cell deficit of Section 6.1) and the electron current 0.205 against 0.213. The hybrids float at $\chi_f = 1.50$, 1.51 and 1.525 and collect 2.19, 2.15 and 2.10 $\Gamma_{th,i}$: the current approaches run A from above and the potential from below as $\delta$ is lowered, and at $\delta = 0.02$ both are within 2–3 percent of run A. The potential is smooth across the interface, where the ion density steps by 0.8 percent at $\delta = 0.05$. Beyond about 1300 time units, no run is steady. A standing oscillation of the bulk potential, with a period of about 170 time units (the ion transit time of the domain), grows in every run, the all-kinetic run A included, and the coarser the threshold, the more the hybrid amplifies it (the $\delta = 0.1$ run diverged at 2300 time units). The mode therefore belongs to the floating configuration on $R_{out} = 12R_p$ and not to the coupling, and the comparison of the hybrids holds on the quasi-steady window only. Restarted at the prescribed bias $\chi_p = 1.545$, run A and the $\delta = 0.05$ hybrid stop growing, so the mode needs the floating feedback. The ion octree holds 33, 17 and 6 percent fewer cells than in run A, and the wall time per ion transit is reduced by at most 11 percent: the octree beyond the interface is masked, but the scan of the leaves for the sensors and the overwrite of the buffer shells cost nearly as much as the cells saved.

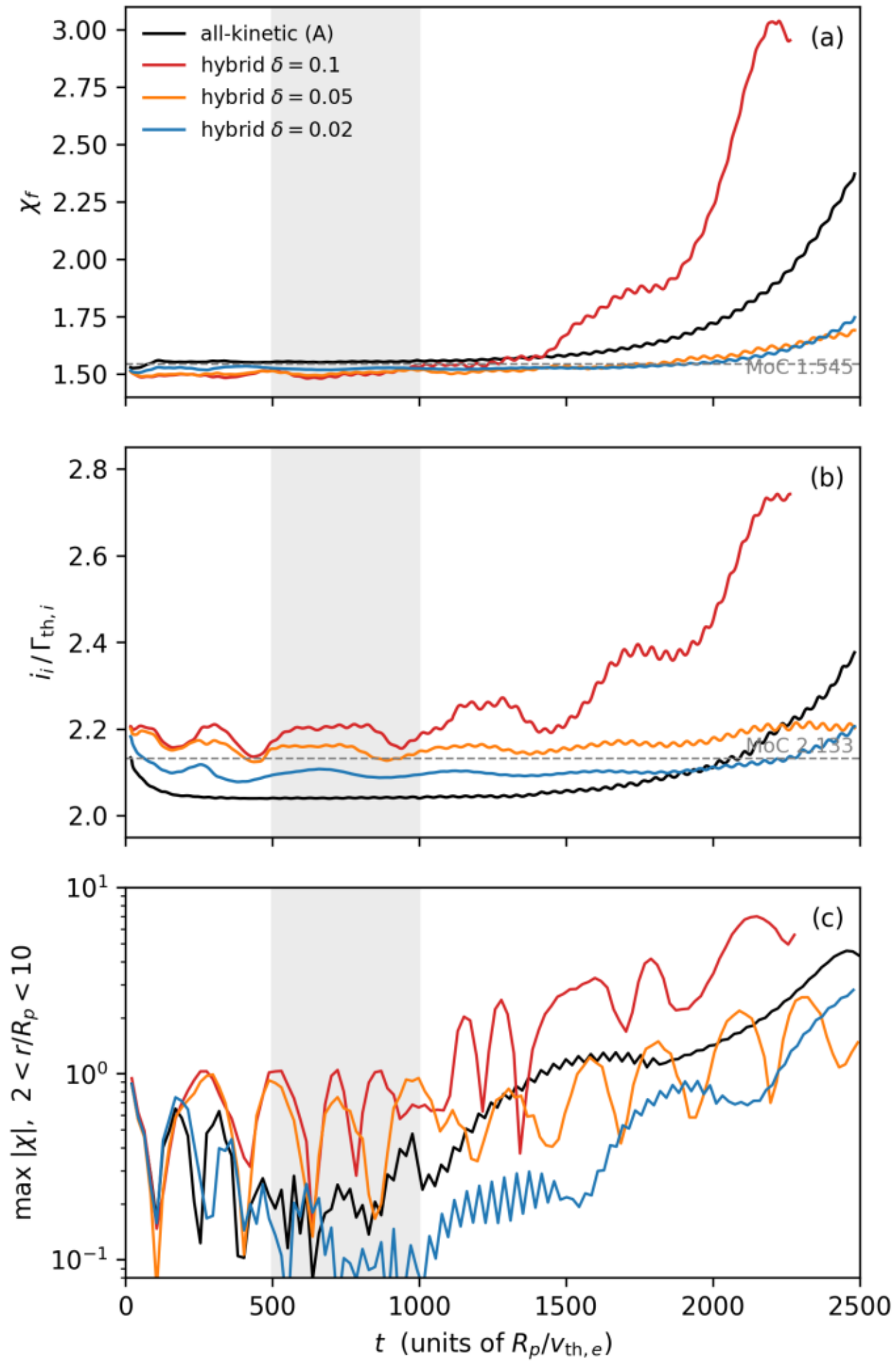

Figure 14. The lock-step ion-hybrid series on the floating screened sphere ($R_p/\lambda_{De} = 8$, $T_i = T_e$, $\chi_f = 1.545$, $R_{out} = 12R_p$): the all-kinetic run A and the hybrids at $\delta = 0.1$, 0.05 and 0.02. (a) Floating potential and (b) collected ion current from the face-flux audit, running means over 30 time units; (c) amplitude of the bulk potential, max$|\chi|$ over $2 < r < 10R_p$, from the radial profiles. Dashed: the MoC values; shaded: the quasi-steady window of the comparison; the $\delta = 0.1$ run was stopped at 2280 time units.

**Charge exchange.** On $R_{out} = 3R_p$, with the free-running clocks under which the all-kinetic series of [1] converged, the hybrid at $\delta = 0.05$ keeps the whole domain kinetic for collisionless ions and at $\lambda_i/R_p = 3$: the anisotropy exceeds the threshold out to $R_{out}$, and the runs reproduce the floating potential and the ion current of the all-kinetic runs to 0.7 percent. At $\lambda_i/R_p = 1$, the interface settles at $2.4R_p$, and the floating potential and the ion current are within 0.4 and 2.7 percent of the all-kinetic run. At $\lambda_i/R_p = 0.3$, charge exchange isotropizes the ions down to the sheath, and the interface moves inward during the run, to $1.6R_p$ after one ion transit and to $1.07R_p$ by the end (Figure 15(a)). There the hybrid fails: it floats at $\chi_f = 2.29$ against 2.67 and collects 0.96 against 0.64 (the all-kinetic run with the same code version; 0.66 in Table 3) $\Gamma_{th,i}$, two thirds of the way to the fluid value 1.11. Lowering the threshold to 0.02 and 0.01 holds the interface at 1.8 and $2.1R_p$ and reduces the excess current to 23 and 13 percent (Figure 15(b, c)), but the ion octree is then as large as in the all-kinetic run. Between the sheath and $2R_p$, the ions are nearly isotropic but are not the drifting Maxwellian of the fluid. The likely cause is the trapped ions, which charge exchange supplies and whose space charge deepens the presheath; the pressure anisotropy does not register them. These runs select the interface by the anisotropy alone, so in the collisional regime the trapping sensor is not optional. None of them saves cells, since on $R_{out} = 3R_p$ the kinetic region either fills the domain or must be kept nearly as large to hold the error near 10 percent.

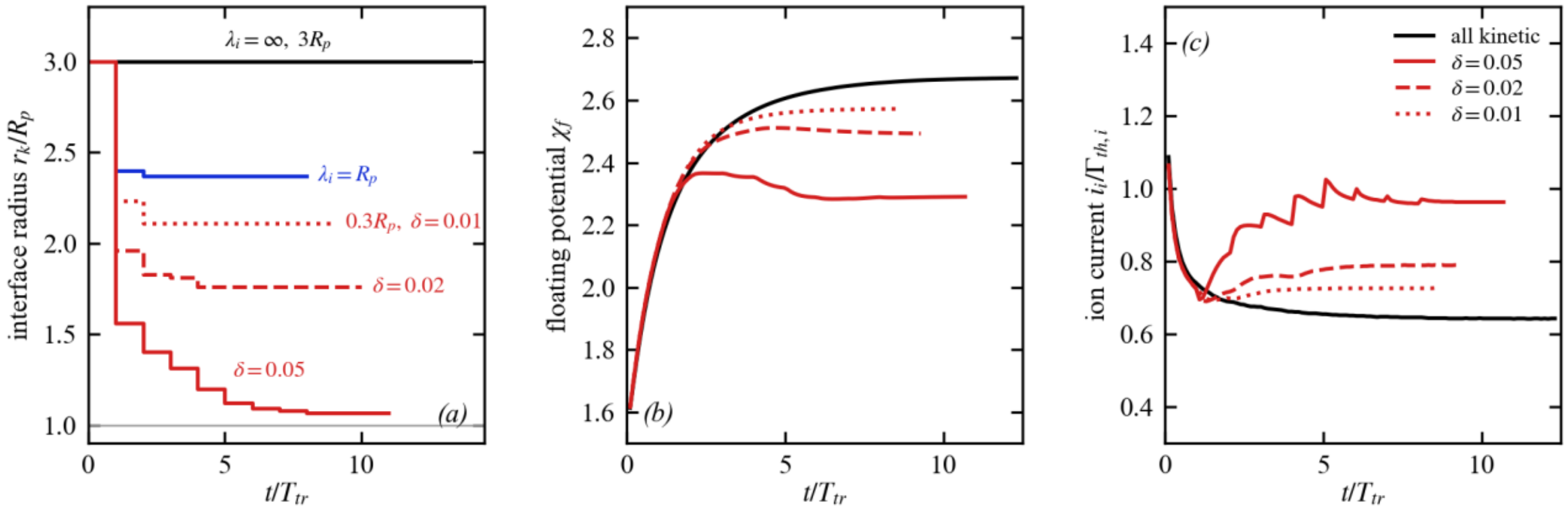

Figure 15. The moving interface of the fluid–1d2v ion hybrid with charge exchange (floating screened sphere, $R_{out} = 3R_p$; ownership re-evaluated every ion transit, $T_{tr} = 40$ time units). (a) Interface radius against time at $\delta = 0.05$ for collisionless ions and $\lambda_i = 3R_p$ (black: the whole domain stays kinetic), $\lambda_i = R_p$ (blue) and $\lambda_i = 0.3R_p$ (red), the last also at $\delta = 0.02$ (dashed) and 0.01 (dotted); grey: the surface. (b) Floating potential and (c) collected ion current at $\lambda_i = 0.3R_p$: the all-kinetic run (black) and the three hybrids.

**Fluid–1d1v(S).** The two-cone model of Section 3 can replace the octree inside, at a fraction of its cost, if its outer boundary condition is generalized. At present, it takes the reservoir Maxwellian at $R_{out}$ and propagates it inward along the orbits. Coupled to a fluid region, it must take the fluid's drifting Maxwellian at the interface, expressed in the invariants $(\varepsilon, \ell^2)$, and propagate that. Everything else, the barrier functions on the potential out to $R_{out}$ and the trapped bath fed by charge exchange, is unchanged. This hybrid has no velocity mesh at all, and its verification against fluid–1d2v on the same cases measures the closure where it matters, at the interface. Both ion hybrids run behind the unchanged solver interface, so the electron solver and the field solver are not touched. The two models are complementary. The two-cone model carries the angular momentum and the trapped bath, for want of which the fluid–1d2v hybrid fails at $\lambda_i = 0.3R_p$, but it has no collisional plasma around it (Section 6.4), and the fluid supplies exactly that. We therefore regard this pair, not the octree, as the ion model for hybrid simulations with PAMAR wherever the orbits are integrable.

## 9.5. Coupling

Both solvers of a hybrid solver cover the whole radial domain. In shells owned by the reduced model, the kinetic distribution is overwritten every step by the state the reduced model implies, and the octree there is masked rather than coarsened (Section 9.4), and the kinetic region receives its inflow through ordinary advection. For the ions, that state is the drifting Maxwellian $f = n(\pi v_{th}^2)^{-3/2}\exp[-(v^2 - 2v\mu u +$ $u^2)/v_{th}^2]$ built from the fluid $n$ and $u$. For collisionless electrons, it is the Liouville state of the current potential, $f_\infty(\varepsilon)$ on the arriving and reflected orbits and zero in the loss cone, which is exact for a repelled species and costs one MoC evaluation. In cells owned by the kinetic solver, the reduced arrays are overwritten by the kinetic moments. Ownership is re-evaluated every ion transit and kept contiguous to the wall, so each species has one interface.

For collisional electrons, the reduced model outside the loss-cone layer is the Fokker–Planck model, which couples to the octree through partial currents per velocity row rather than by overwriting cells (Section 9.3).

The saving of the ion hybrid is modest in 1d2v. On the logarithmic radial map, the ion kinetic region $5.2R_p$ holds about 67 percent of the radial cells and the electron region $2.1R_p$ about 30 percent, while the electron octree, the expensive one, shrinks 1.6 to 5.3 times in the measured electron hybrid [1]. These demonstrations test the coupling and the sensors rather than speed. The economy grows with the domain (the OML approach needs $R_{out} \gg R_p$), with the collisional time scale over which charge exchange fills the trapped island, and with the phase-space dimension in problems without spherical symmetry.

## 10. Discussion and conclusions

We studied the absorbing sphere in plasma with PAMAR, using independent kinetic, reduced, and fluid solvers for electrons and ions, and exact references for the collisionless and elastically scattering problems. We obtained the current–voltage characteristics of a spherical probe and analyzed the potential, densities, drift velocities, and temperatures of the two species in collisionless and collisional regimes. Far from the sphere, the potential follows the Coulomb law in neither limit: it falls as $r^{-2}$ without collisions and as $r^{-1}$ only beyond the ion mean free path. With elastic electron scattering, the 1d2v solver follows the exact reduction of the collected current with the mean free path, while its absolute current carries the wall-cell deficit of the mesh.

Our comparisons of the ion solvers show that the fluid closure fails around a sphere because ions arrive with angular momentum, and that the failure is confined to a few probe radii whose extent the pressure anisotropy measures. The two-cone model carries the angular momentum analytically and reproduces the floating potential and the collected current at a fraction of the octree's cost. With charge exchange, it follows the kinetic current within 7 percent down to $\lambda_i = R_p$, but it has no collisional plasma far from the sphere, which a fluid model outside must supply. That is the premise of algorithm refinement, which we have demonstrated in Section 9 and in the companion paper [1]. For the electrons, we keep the 1d2v solver within about a mean free path of the body, where the loss cone makes the escaping electrons anisotropic; everywhere else, we use a Fokker–Planck model in position and total energy. For the ions, a fluid model outside is coupled to a kinetic solver, and the two are selected by the pressure anisotropy. With charge exchange at $\lambda_i = 0.3R_p$, the anisotropy alone hands the nearly isotropic ions near the probe to the fluid, which then over-collects by 50 percent, and the trapping sensor is needed to place the interface. The electron layer and the ion orbital region emerge at different radii, which supports selection per species and per cell. Future work will focus on the trapping sensor and on the fluid–1d1v(S) ion hybrid.

## Acknowledgments

This work was supported by the NSF EPSCoR FTPP project (Cooperative Agreement OIA-2148653).

## Declaration of competing interests

The authors declare no known competing financial interests or personal relationships that could have influenced the work reported in this paper.

## Data availability

Data will be made available on request.

## Declaration of generative AI and AI-assisted technologies in the manuscript preparation process

During the preparation of this work, the authors used Claude Code (Anthropic) to assist with code development, data analysis, and manuscript editing. After using this tool, the authors reviewed and edited the content as needed and take full responsibility for the published article.